\documentclass[preprint,twocolumn]{aastex}
\usepackage{graphicx, subfigure}
\usepackage{amssymb}
\usepackage{amsmath}
\usepackage{epstopdf}
\usepackage{xcolor}
\usepackage{natbib}
\usepackage{enumerate}
\usepackage{multirow}
\usepackage{xspace}
\usepackage{url}
\usepackage[normalem]{ulem}
\usepackage[yyyymmdd,hhmmss]{datetime}

\newcommand{\Fermi}{\emph{Fermi}\xspace}

\usepackage{amsmath,amstext,natbib}
\usepackage[colorlinks=false,pdfborder={0 0 0},hyperfootnotes=true]{hyperref}
\usepackage[switch]{lineno}

\shorttitle{\Fermi Observations of the LIGO Event from O1}
\shortauthors{\Fermi GBM \& LAT Collaborations}

\begin{document}
\onecolumn
\title{Searching the Gamma-ray Sky for Counterparts to Gravitational Wave Sources: \\
\Fermi GBM and LAT Observations of LVT151012 and GW151226}
\author{
J.~L.~Racusin\altaffilmark{1*},
E.~Burns\altaffilmark{2*},
A.~Goldstein\altaffilmark{3*},
V.~Connaughton\altaffilmark{3},
C.~A.~Wilson-Hodge\altaffilmark{4},
P.~Jenke\altaffilmark{5},
L.~Blackburn\altaffilmark{6},
M.~S.~Briggs\altaffilmark{5,7},
J.~Broida\altaffilmark{8},
J.~Camp\altaffilmark{1}
% ...
\vspace{5cm}
}
\altaffiltext{*}{Corresponding authors: J.~L.~Racusin, judith.racusin@nasa.gov;  E.~ Burns, EricKayserBurns@gmail.com; A. Goldstein, adam.m.goldstein@nasa.gov;  N.~Omodei, nicola.omodei@stanford.edu;  G.~Vianello, giacomov@stanford.edu.}
% ...
\altaffiltext{1}{NASA Goddard Space Flight Center, Greenbelt, MD 20771, USA}
\altaffiltext{2}{Physics Dept, University of Alabama in Huntsville, 320 Sparkman Dr., Huntsville, AL 35805, USA}
\altaffiltext{3}{Universities Space Research Association, 320 Sparkman Dr. Huntsville, AL 35806, USA}
\altaffiltext{4}{Astrophysics Office, ZP12, NASA/Marshall Space Flight Center, Huntsville, AL 35812, USA}
\altaffiltext{5}{CSPAR, University of Alabama in Huntsville, 320 Sparkman Dr., Huntsville, AL 35805, USA}
\altaffiltext{6}{LIGO, Massachusetts Institute of Technology, Cambridge, MA 02139, USA}
\altaffiltext{7}{Dept. of Space Science, University of Alabama in Huntsville, 320 Sparkman Dr., Huntsville, AL 35805, USA}
\altaffiltext{8}{Physics and Astronomy, Carleton College, MN, USA 55057}
% ...

\begin{abstract}
We present the \Fermi Gamma-ray Burst Monitor (GBM) and Large Area Telescope (LAT) observations of the LIGO binary black hole merger event GW151226 and candidate LVT151012.  No candidate electromagnetic counterparts were detected by either the GBM or LAT.  We present a detailed analysis of the GBM and LAT data over a range of timescales from seconds to years, using automated pipelines and new techniques for characterizing the upper limits across a large area of the sky. Due to the partial GBM and LAT coverage of the large LIGO localization regions at the trigger times for both events, differences in source distances and masses, as well as the uncertain degree to which emission from these sources could be beamed, these non-detections cannot be used to constrain the variety of theoretical models recently applied to explain the candidate GBM counterpart to GW150914.

\end{abstract}

\maketitle
\twocolumn

\section{Introduction}
The era of multi-messenger astronomy has fully begun with the regular detections of gravitational waves (GWs) from merging compact objects by the Laser Interferometer Gravitational-wave Observatory (LIGO; \citealt{2016PhRvL.116f1102A}), and large multi-wavelength campaigns to pursue electromagnetic (EM) counterparts \citep{2016arXiv160208492A}.  As demonstrated with GW150914, \Fermi's GBM and LAT are uniquely capable of providing all-sky observations from hard X-ray to high-energy $\gamma$-rays in normal survey operations, including covering the entire localization probability maps of LIGO events \citep{GW150914_GBM,GW150914_LAT,2016arXiv160208492A} within hours of their detections (see also \citealt{GW150914_AGILE}).

In addition to GW150914 \citep{2016PhRvL.116f1102A,2016arXiv160203839T}, two other candidate compact object merger events were reported by LIGO during the O1 observing run from 2015 September 12 to 2016 January 12.  
GW151226 and the sub-threshold LIGO-Virgo Trigger LVT151012 (if the latter is from a real astrophysical event) are associated with the mergers of two compact objects, likely both stellar-mass black holes (BHs) \citep{GW151226_LIGO}. 

Prior to the watershed discovery of GWs from the binary black hole (BBH) merger GW150914, and the candidate $\sim 1$ s long $\gamma$-ray counterpart GW150914-GBM that was seen 0.4 s later \citep{GW150914_GBM}, there was little theoretical expectation for EM counterparts to BBH mergers.  The weak $\gamma$-ray signal observed by the GBM is temporally and spatially coincident with the GW trigger, and appears similar to a low-fluence short Gamma-ray Burst (sGRB).  Note that the candidate GBM counterpart was not detected by the INTEGRAL SPI-ACS (Anti-Coincidence Shield; \citealt{GW150914_integral}), and there is debate regarding the nature of the GBM signal \citep{greiner16}.
Since the potential discovery was announced, innovative ideas have emerged to explain an observational signature that possibly resembles a weak sGRB from a BBH (e.g., \citealt{Loeb2016}, \citealt{Fraschetti2016}, \citealt{Janiuk2016}, and \citealt{2016ApJ...821L..18P}); see also \citet{Lyutikov2016} for significant constraints on such models.  
Binary Neutron Star (BNS) or Neutron Star - Black Hole (NS-BH) mergers are the most likely progenitors of sGRBs (\citealt{Eichler1989}, \citealt{Narayan1992}, \citealt{LeeRamirezRuiz2007}, and \citealt{Nakar2007}), and therefore they are the most similar object class for comparison to \Fermi observations of BBH mergers.

Approximately 68\% (for GBM) and 47\% (for LAT) of the LVT151012 LIGO localization probability and 83\% (for GBM) and 32\% (for LAT) of the GW151226 LIGO localization probability were within the \Fermi GBM and LAT fields of view (FoV) at the trigger times, respectively.
The GBM and LAT completed their first post-trigger coverage of the entire localization probability map for LVT151012 within 8 minutes (for GBM) and 113 minutes (for LAT), and for GW151226 within 34 minutes (for GBM) and 140 minutes (for LAT).

No credible counterpart candidates were detected by either the GBM or the LAT at the trigger times of both events or on the timescales of minutes, hours, days, and months afterwards.  These non-detections do not constrain models proposed for 
the candidate GBM counterpart to GW150914, owing to the partial GBM and LAT coverage of the LIGO localization region at the time of trigger for both events, differences in the source distances and system masses, as well the uncertain degree to which emission from these sources could be beamed.  Therefore, these GBM and LAT non-detections do not provide strong evidence whether $\gamma$-ray emission is associated with BBH mergers.

A statistically-significant sample of BBH mergers, which will be collected over the coming years by the advanced network of GW observatories (including LIGO and Virgo) and wide-field $\gamma$-ray instruments, will be required to understand the nature of candidate EM counterparts to BBH merger events, such as GW150914-GBM.

A summary of the pertinent information regarding the LIGO sources is provided in Section \ref{sec:ligo}, and the custom data analysis and results of specialized searches for $\gamma$-ray counterparts are discussed in Section \ref{sec:gbm_analysis} (GBM) and Section \ref{sec:lat_analysis} (LAT).  In Section \ref{sec:disc}, we discuss the implications of these non-detections on counterpart searches in general and specifically for GW150914-GBM, placing our GW counterpart limits in the context of sGRB properties. We further comment on the relevance of these observations to the recent theoretical developments regarding how a $\gamma$-ray counterpart might be produced by a BBH merger. Finally, we conclude in Section \ref{sec:conc}.

\section{Observations and Data Analysis}

This section describes several standard and new extensive searches of the GBM and LAT data within the LIGO localization contours of LVT151012 and GW151226 using a variety of techniques and timescales.  The timescales referred to throughout this section are summarized in Table \ref{table:timescales}.  There were no credible counterpart candidates detected in any of these searches.


\begin{table*}
\centering
\small
\begin{tabular}{|c|c|cc|}
\cline{3-4}
\multicolumn{2}{c|}{} & \multicolumn{2}{c|}{\bf Analysis Time Period} \\
\cline{3-4}
\multicolumn{2}{c|}{\ } & {\bf LVT151012} & {\bf GW151226} \\
\hline
\multirow{3}{*}{\rotatebox[origin=c]{90}{GBM}} & $T_{\rm blind}$ & continuous & continuous \\
& $T_{\rm seeded}$ & $\pm$30 s & $\pm$30 s \\
& $T_{\rm EOT}$ & 1 day, 1 month, 1 year$^{\#}$ & 1 day, 1 month, 1 year$^{\#}$ \\
\hline
\multirow{6}{*}{\rotatebox[origin=c]{90}{LAT}} & $T_{\rm fixed_1}$ & $-$10 -- +10 s & $-$10 -- +10 s \\
& $T_{\rm fixed_2}$ & 0--8 ks & 0--1.2 ks \\
& $T_{\rm fixed_3}$ & -- & 0--10 ks  \\
& $T_{\rm adaptive}$ & 130--4500 s$^{*}$ & 350--2900 s$^{*}$ \\
& $T_{\rm ASP}$ & 6 hours, 1 day & 6 hours, 1 day \\
& $T_{\rm FAVA}$ & $\pm$1 week & $\pm$1 week \\
\hline
\end{tabular}
\caption{Timescales over which the GBM and LAT data were studied with the various analyses of LVT151012 and GW151226, discussed in Sections \ref{sec:gbm_analysis} \& \ref{sec:lat_analysis}, all referenced to the LIGO trigger times ($t_{\rm LVT}$ or $t_{\rm GW}$). $^{*}$Note that for $T_{\rm adaptive}$ we report the minimum and maximum possible duration. $^{\#}$Note that $T_{\rm EOT}$ straddles $t_{\rm LVT}$ and $t_{\rm GW}$ as evenly as possible given limitations of when this analysis was performed relative to the triggers.}
\label{table:timescales}
\end{table*}

\subsection{LIGO}\label{sec:ligo}

LVT151012 was detected at both the LIGO Hanford and Livingston facilities using the offline data analysis pipelines \texttt{gstlal} \citep{2016arXiv160404324M} and \texttt{pycbc} \citep{2015arXiv150802357U}, designed to detect compact binary coalescence (CBC) events, with the candidate source being detected at 09:54:43.4 UTC on 2015 October 12 (hereafter $t_{\rm LVT}$), with $\sim 2 \sigma$ significance.  The LIGO GW analysis of LVT151012 yields a relatively high false alarm rate (FAR) of 1 per 2.3 years, BH masses of $23^{+18}_{-6}$ and $13^{+4}_{-5}$ M$_{\odot}$, and a distance of $1100\pm500$ Mpc \citep{2016arXiv160203839T}.

GW151226 was detected at both the LIGO Hanford and Livingston facilities, using the \texttt{gstlal} CBC real-time pipeline, at 03:38:53.6 UTC on 2015 December 26  (hereafter $t_{\rm GW}$).  The GW analysis provides a FAR of less than 1 per 1000 years, and parameter estimation provides BH masses of $14.2^{+8.3}_{-3.7}$ and $7.5\pm 2.3$ M$_{\odot}$, and a distance of $440^{+180}_{-190}$ Mpc \citep{GW151226_LIGO}.

\begin{figure}[t!]
\centering
\includegraphics[width=0.48\textwidth,trim=190 50 190 50,clip=true]{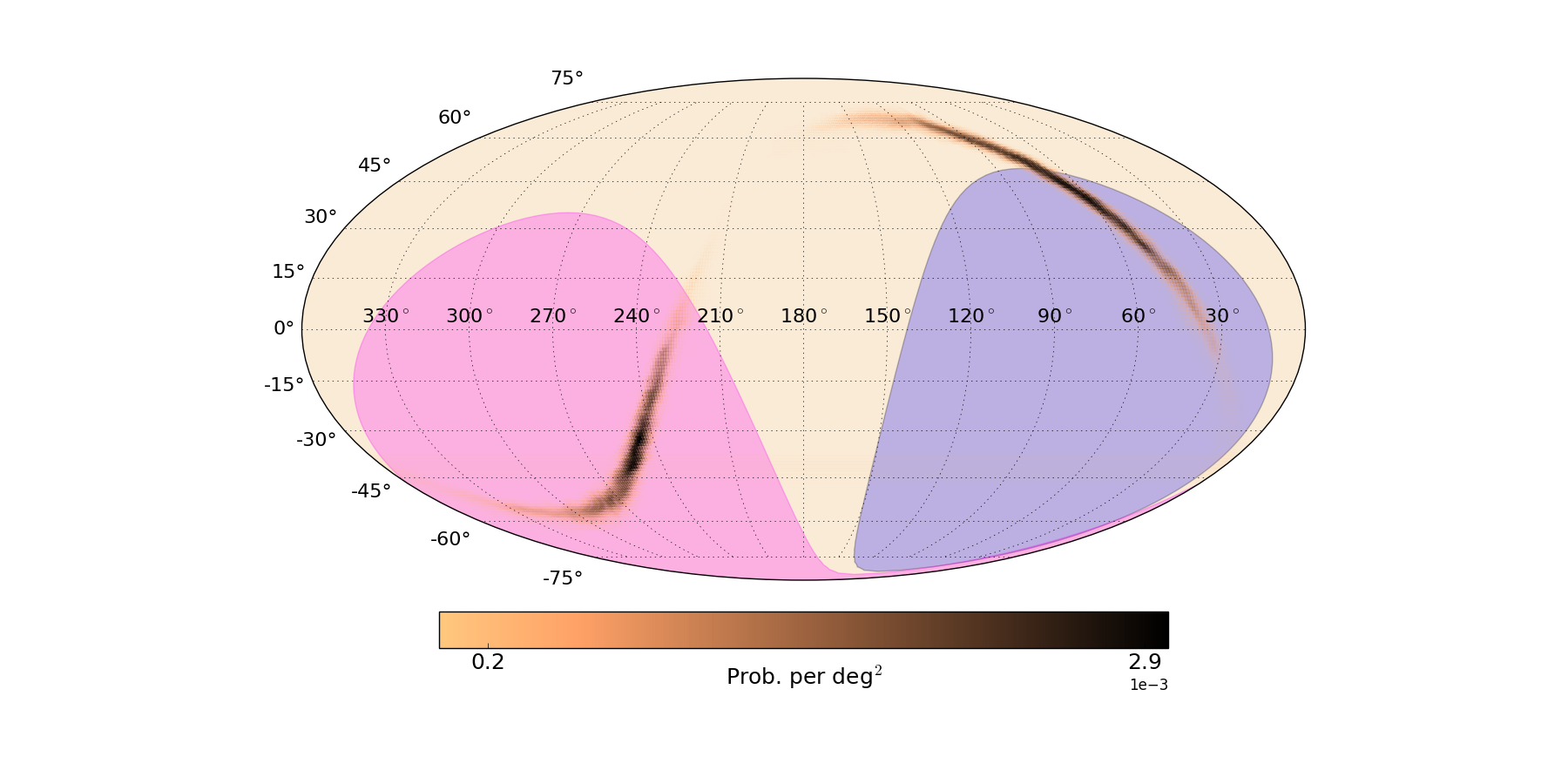}
\includegraphics[width=0.48\textwidth,trim=190 50 190 50,clip=true]{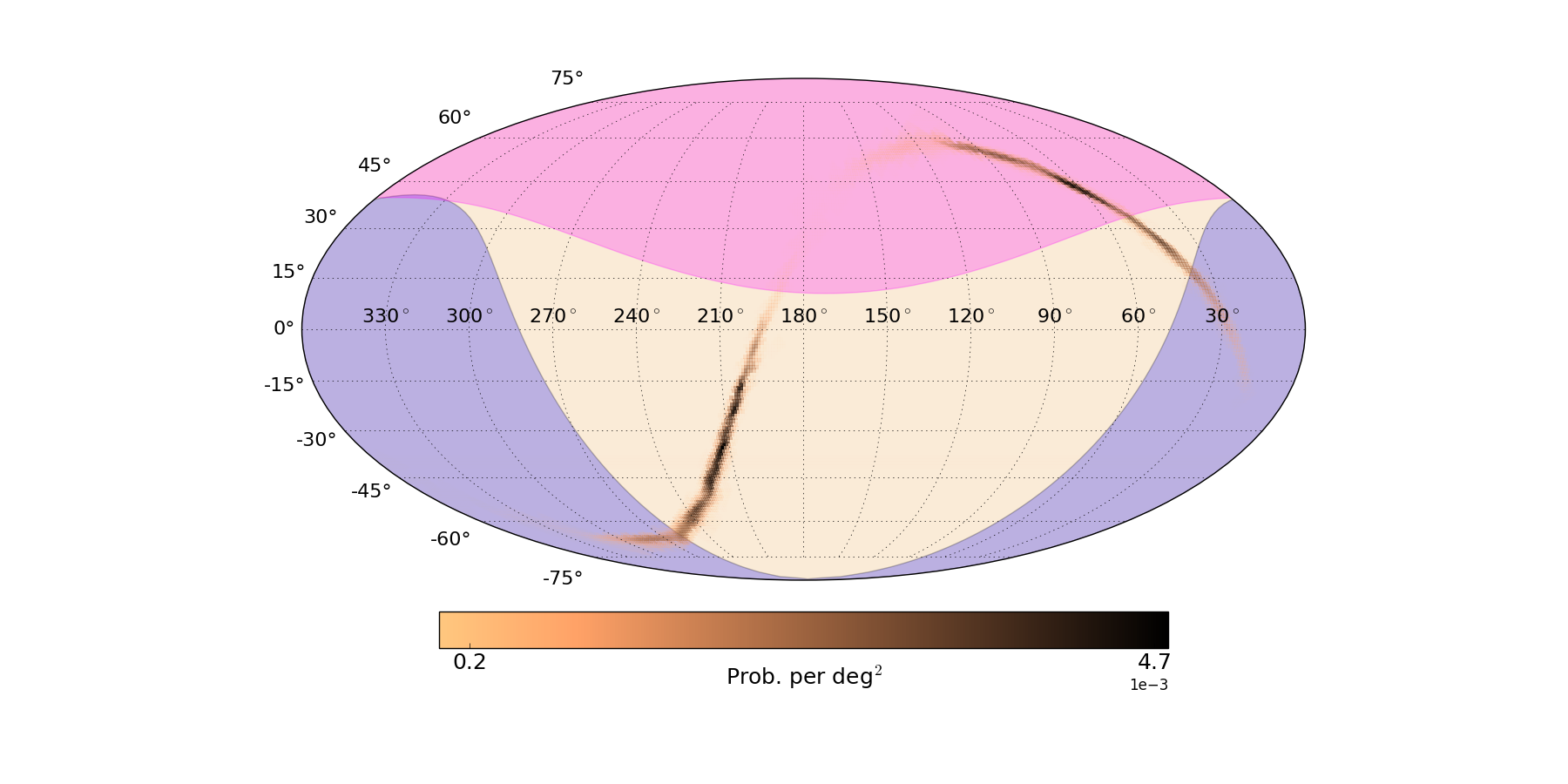}
\caption{LIGO localization probability maps for LVT151012 ({\it top}; \citealt{LVT151012_GCN}) and GW151226 ({\it bottom}; \citealt{GW151226_GCN}) indicating the portions of the sky occulted by the Earth for \Fermi at the time of the LIGO trigger (blue shaded region). The GBM observes the entire unocculted sky. The pink shaded region indicates the portions of the sky within the LAT FoV at the GW trigger times.} \label{fig:ligomaps}
\end{figure}

The LIGO Scientific Collaboration and the Virgo Collaboration reported the discovery and results from Bayesian parameter estimation analyses of LVT151012 and GW151226 under the assumption that the signals arise from a CBC using the latest offline calibration of the GW strain data (\citealt{2016arXiv160203839T}, \citealt{GW151226_LIGO}). The most accurate localization maps for these events (LALInference, \citealt{2015PhRvD..91d2003V}) are based on Bayesian Markov-Chain Monte Carlo and nested sampling to forward model the full GW signal including spin precession and regression of systematic calibration errors. The analysis of the \Fermi observations requires only the trigger times and localization maps as inputs, which were provided via the Gamma-ray Coordinates Network (GCN; \citealt{LVT151012_GCN,GW151226_GCN}) to groups with a memorandum of understanding with LIGO.  The LIGO localization maps for LVT151012 and GW151226 are shown in Figure \ref{fig:ligomaps} with the regions occulted by the Earth for \Fermi at the times of the GW triggers, indicating the portions of the sky and LIGO localization probability regions visible to both the GBM and LAT.  All of the GBM and LAT upper limit measurements are calculated for the LIGO localization regions containing 90\% of the probability.  The following sections provide further details on the GBM and LAT observations and analyses.

\subsection{GBM}\label{sec:gbm_analysis}
The GBM is composed of 12 Sodium Iodide (NaI) detectors and two Bismuth Germanate (BGO) detectors \citep{2009ApJ...702..791M}, with the NaIs providing sensitivity between 8 keV and 1~MeV (NaIs), and the BGOs extending the energy range to 40 MeV. The detectors are spaced around the \Fermi spacecraft, oriented at different angles to provide approximately uniform sky coverage, resulting in an instantaneous FoV of $\sim$70\% of the sky, with the remainder blocked by the Earth. The GBM operates continuously except during passages through the South Atlantic Anomaly (SAA), reducing the time-averaged sky exposure to $\sim$60\%. The data types relevant to the analyses in this paper are CTIME, which is binned at 0.256 s intervals into 8 energy channels, and continuous time-tagged event (CTTE) which is unbinned in time and in 128 energy channels. For more detailed explanations of the GBM instrumentation, data types, the triggering algorithms, the sub-threshold searches, and the persistent source searches see \cite{GW150914_GBM} and the respective papers for each technique  (\citealt{2009ApJ...702..791M,2015ApJS..217....8B,2012ApJS..201...33W}; \Fermi-GBM Collaboration 2016, in preparation).

The GBM triggers on board in response to impulsive events when the count rates recorded in two or more NaI detectors significantly exceed the background count rate on at least one timescale (from 16 ms to 4.096 s) in at least one of three energy ranges above 50 keV (50--300 keV, $>$100 keV, $>$300 keV). The GBM also triggers on softer events (25--50 keV) on shorter timescales (from 16 to 128 ms). Since November 2009 the GBM also triggers on significant increases above the background count rate in the BGOs.

As described in \cite{GW150914_GBM}, two new GBM ground pipelines are designed to maximize the chances of detecting counterparts to GW events while carefully accounting for fluctuations common in a background-dominated measurement.  The GBM offline blind-search pipeline\footnote{\url{http://gammaray.nsstc.nasa.gov/gbm/science/sgrb_search.html}} (\Fermi-GBM Collaboration 2016, in preparation) is sensitive to impulsive transients too weak to trigger on board. The pipeline searches CTTE data over 0.1--2.8 s timescales and in four energy bands spanning $\sim$30--1000 keV, approximately doubling the sensitivity of the GBM to sGRBs.  
The GBM seeded-search pipeline \citep{2015ApJS..217....8B} uses the GW trigger time and (optionally) the sky location to inform a maximum-likelihood search for modeled burst signals in the GBM data (assuming one of three template source spectra). Using an existing catalog of the GBM all-sky instrumental response models \citep{2015ApJS..216...32C}, the search procedure is to calculate expected source counts for each detector, and compare this predicted signal to any observed excess detector counts over background. An overlapping set of short foreground intervals between 0.256 and 8 s long is tested for the contributions from a modeled burst, covering a total search interval of $\pm$30s ($T_{\rm seeded}$) about the GW trigger time.  The seeded-search pipeline combines NaI and BGO data to provide a
sensitive search for short-duration transients.   This search will be expanded
in the future to use the significance of a sub-threshold signal in
either the GBM or GWs to strengthen the detection of a signal in the
other, provided the false positive rate of the joint search is
characterized and the detection levels in both instruments are
selected accordingly.  The ability to validate sub-threshold 
candidates effectively boosts the LIGO/Virgo horizon by 15-20\% and
thus the search volume by 50-75\% \citep{2013PhRvD..87l3004K,2015ApJS..217....8B}.  

In the absence of a detected counterpart signal, we have developed a new technique for setting limits on the strength of impulsive $\gamma$-ray emission. The LIGO probability map is divided into regions best observed by the same NaI detector. A 3$\sigma$ upper limit on the count rate is defined as three times the standard deviation around a background fit that excludes $\pm$30 s from the GW trigger time. This can be converted to a flux upper limit by taking the counts and folding an assumed model through the response. We assume a cutoff power-law fit with $E_{peak}$ = 566~keV and a photon index of 0.42, which are the values at peak density for sGRBs best fit by a cutoff power-law from the GBM spectral catalog\footnote{\url{http://heasarc.gsfc.nasa.gov/W3Browse/fermi/fermigbrst.html}} \citep{gruber14,goldstein12} after accounting for parameter correlation. With an assumed distance, these upper limits can be converted to luminosity upper limits.

In addition to searching for impulsive events, the GBM can act as an all-sky monitor for hard X-ray sources over longer timescales using the Earth Occultation Technique (EOT; \citealt{2012ApJS..201...33W}). The EOT stacks the differences in the background count rates as a source sets or rises behind the Earth, and searches the 12--25, 25--50, 50--100, 100--300, and 300--500 keV energy bands.  We applied the EOT over timescales of 1 day before and after the GW trigger date, 1 month starting at the GW trigger date, and 1 year centered as closely to the GW trigger as possible (given limitations of data collected at the time that this analysis was performed). We also now calculate direction-dependent upper limits for persistent emission owing to the extended LIGO localizations.

\subsubsection{GBM Observations of LVT151012}

The GBM collected data continuously, without passing through the SAA, from 24 minutes prior to 50 minutes after the LIGO detection of LVT151012 ($t_{\rm LVT}$). Figure \ref{fig:ligomaps} shows the LIGO sky map from \cite{LVT151012_GCN} with the blue shaded region indicating the region of sky occulted by the Earth for \Fermi at the time of the GW event. The GBM was observing 68.2\% of the LIGO localization probability at $t_{\rm LVT}$, with exposure of the rest of the localization region over the next 8 minutes.

\begin{figure}[t!]
\centering
\includegraphics[width=0.48\textwidth]{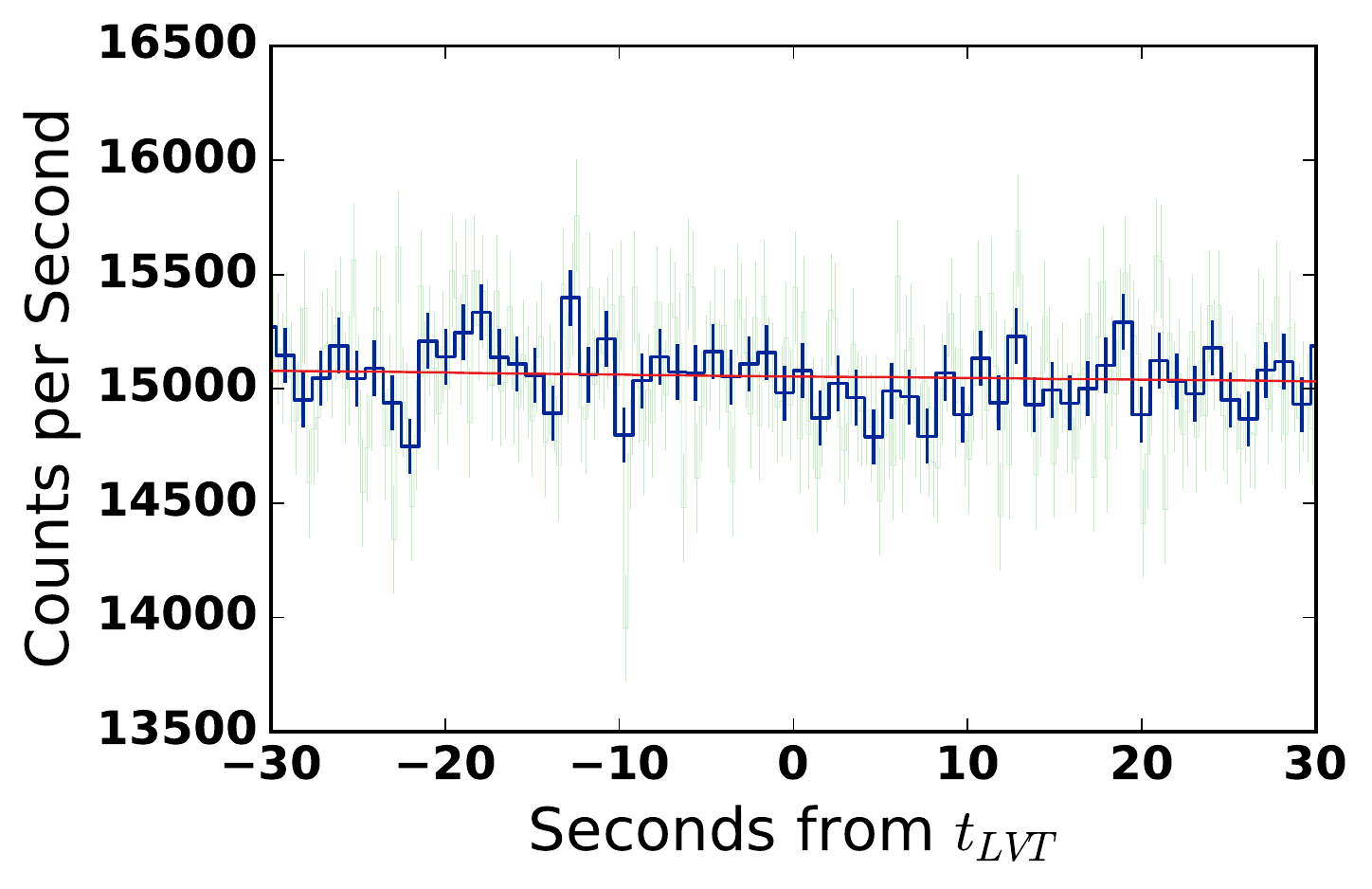}
\caption{There is no evidence that the GBM detected any significant emission during LVT151012, demonstrated by the summed count rate light curve over all GBM detectors (NaI from $\sim$10--1000 keV, BGO from 0.4--40 MeV) during the $T_{\rm seeded}$ interval: $-30<t_{\rm LVT}<+30$ s. The blue curve shows CTTE data rebinned into 1.024 s bins, the green curve is standard CTIME data with 0.256 s bins, and the red is a sum of non-parametric fits of the background of each detector and CTIME energy channel. There are no statistically significant fluctuations within this interval.}
\label{fig:GBMLVT151012summed}
\end{figure}

The only GBM on-board trigger within 12 hours of LVT151012 was misclassified as a GRB by the flight software, and was determined to be caused by a high local particle flux due to an exit from the SAA. The offline blind-search pipeline found no credible candidates within 2 days of the LIGO trigger. There were also no candidates found by lowering the threshold in a 10-minute time window around $t_{\rm LVT}$. The seeded-search pipeline was run on the $T_{\rm seeded}$ interval of $-30<t_{\rm LVT}<+30$ s, searching for a potential counterpart with duration between 0.256 s and 8 s.  The interval was selected {\it a priori} roughly guided by the assumption that if GRBs are related to compact binary mergers then the impulsive $\gamma$-ray emission should be close in time to the GWs, with a wide enough search window to catch possible precursor emission \citep{2010ApJ...723.1711T} and possibly unexpected time offsets from $t_{\rm LVT}$. A light curve showing the summed count rate (ignoring the lowest and highest energy standard CTIME channels) is shown in Figure \ref{fig:GBMLVT151012summed}.

We find no evidence for the counterpart reported by \cite{2016arXiv160306611B} in their search of the GBM data around LVT151012.  Our search method combines signals in the 14 GBM detectors in a way that tests for the likelihood of a source from any sky position.  This is done by weighting both the contribution from each detector and the contribution of each energy channel according to their expected relative contributions for a source at that position.  By using the detector responses rather than examining just the raw count rates above background, we can find weak sources that are consistent with an astrophysical source while rejecting fluctuations of similar magnitude in counts space.  That we do not find the candidate counterpart reported in \cite{2016arXiv160306611B} suggests that either the relative rates among detectors or the distribution of counts in energy for their event are not indicative of a physical source from a single sky position.  Indeed, \cite{2016arXiv160306611B} state they do not use the response information to weight the relative signals when combining detector information, instead weighting the contributions of each detector and energy channel according to signal-to-noise ratio above the background count rates, without consideration as to whether the weighted spectrum is physical or the detector weights are consistent with an arrival direction from a single position. Sub-threshold events in background-limited detectors are weak and each detector energy channel is subject to fluctuations.   The robustness of our technique relies on the combination of 14 individual measurements in a coherent way that uses knowledge of detector responses and typical source energy spectra.

\begin{figure}[t!]
\centering
\includegraphics[width=0.48\textwidth,trim=190 80 190 62,clip=true]{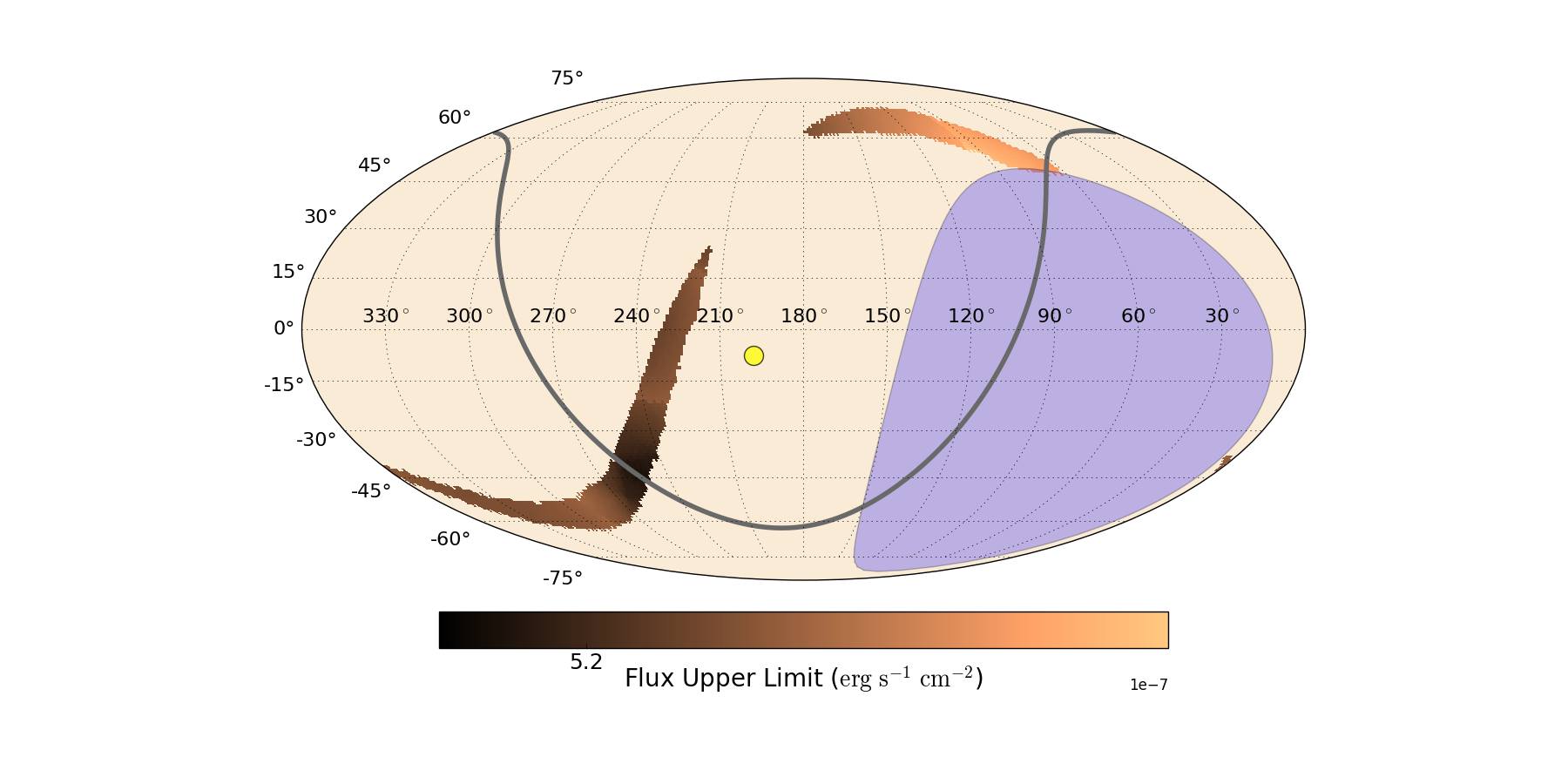}
\caption{The area within the LVT151012 LIGO localization contour is shaded to indicate the GBM 10--1000 keV flux upper limits during during the $T_{\rm seeded}$ interval: $-30<t_{\rm LVT}<+30$ s.  The purple shaded region indicates where the sky was occulted by the Earth for \Fermi.  The Galactic plane is the grey curve, and the Sun is indicated by the yellow disk.}
\label{fig:GBMLVT151012upperlimits}
\end{figure}

Given the lack of any significant impulsive $\gamma$-ray emission above the background, we set upper limits on the impulsive emission (Figure \ref{fig:GBMLVT151012upperlimits}). Using the EOT, we also searched for longer-lasting emission: 1 day before and 1 day after $t_{\rm LVT}$, a month starting at $t_{\rm LVT}$ (2015 October 12 to 2015 November 11), and a year centered around $t_{\rm LVT}$ (2015 April 12 to 2016 April 12). No new sources were detected on any of the searched timescales and energy bands.

\subsubsection{GBM Observations of GW151226}

The GBM collected data, without passing through the SAA, continuously from nearly 30 minutes before to almost 10 hours after GW151226 ($t_{\rm GW}$). Figure \ref{fig:ligomaps} shows the LIGO sky map from \cite{GW151226_LIGO}, and the regions of the sky accessible to the GBM and LAT at the time of detection of the GW event. The GBM observed 83.4\% of the LIGO localization probability during the GW emission of GW151226, with exposure of the rest of the localization region over the next 34 minutes.

\begin{figure}[t!]
\centering
\includegraphics[width=0.5\textwidth]{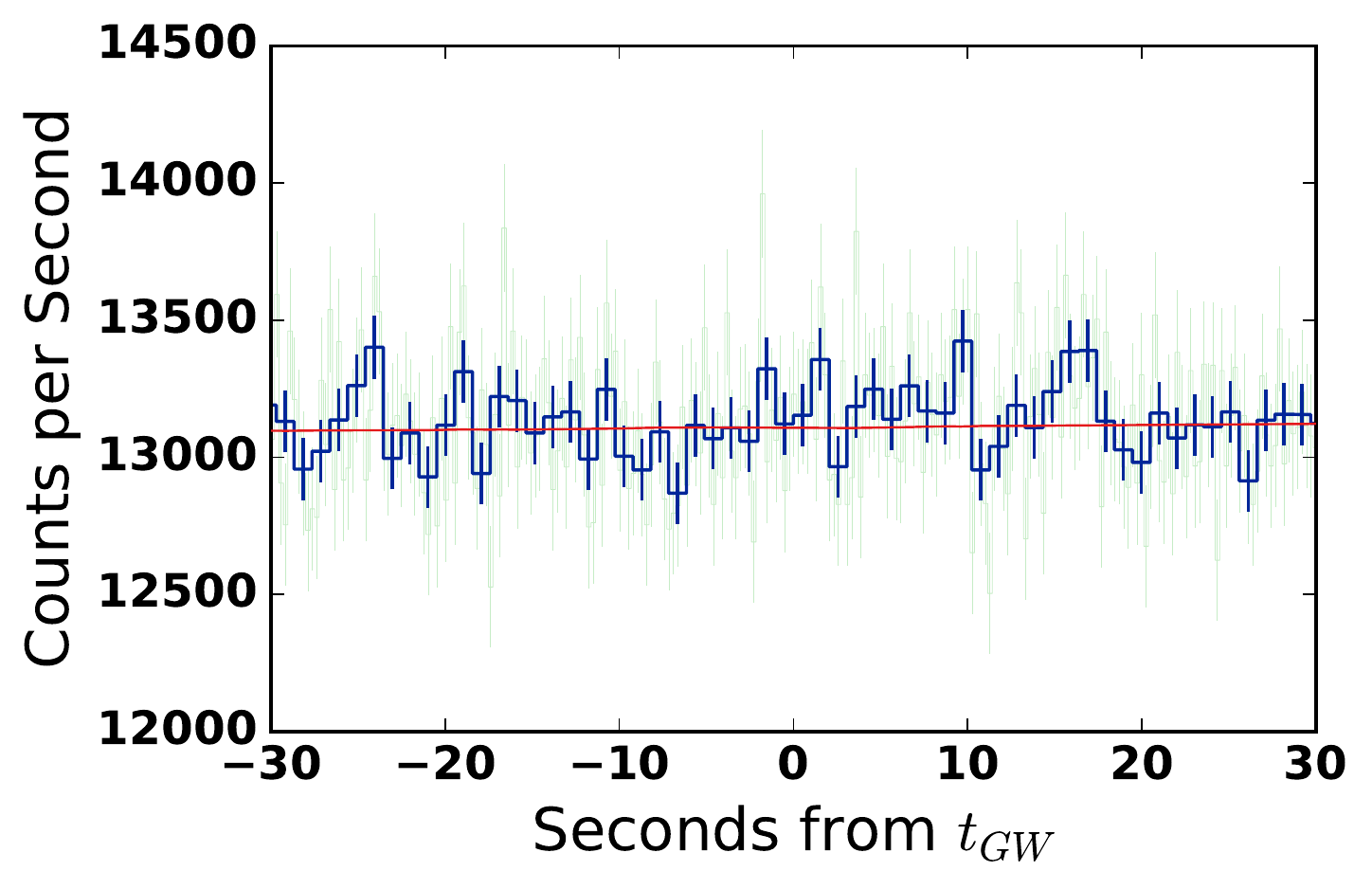}
\caption{There is no evidence that the GBM detected any significant emission during GW151226, demonstrated by the summed count rate light curve over all GBM detectors (NaI from $\sim$10--1000 keV, BGO from 0.4--40 MeV) during the $T_{\rm seeded}$ interval: $-30<t_{\rm GW}<+30$ s.  The blue curve shows CTTE data rebinned into 1.024 s bins, the green curve is standard CTIME data with 0.256 s bins, and the red curve is a sum of a non-parametric fit of the background of each detector and CTIME energy channel. There are no statistically significant fluctuations within this interval.}
\label{fig:GBMGW151226summed}
\end{figure}

There were no GBM on-board triggers within twelve hours of GW151226, and no candidate counterparts found using the blind-search pipeline within 5 days of $t_{\rm GW}$. There were also no candidates found by lowering the threshold in a 10 minute window around $t_{\rm GW}$. The seeded-search pipeline also found no credible candidates in the $\pm$30 s $T_{\rm seeded}$ interval. The most significant fluctuation identified has a FAR value of $2.2\times 10^{-3}$ and occurred 2.0 s before GW151226. The post-trials False Alarm Probability (FAP) is 20\%; this event is insignificant. A summed count rate light curve (ignoring the lowest and highest energy standard CTIME channels) is shown in Figure \ref{fig:GBMGW151226summed}.

\begin{figure}[t!]
\centering
\includegraphics[width=0.5\textwidth,trim=190 80 190 62,clip=true]{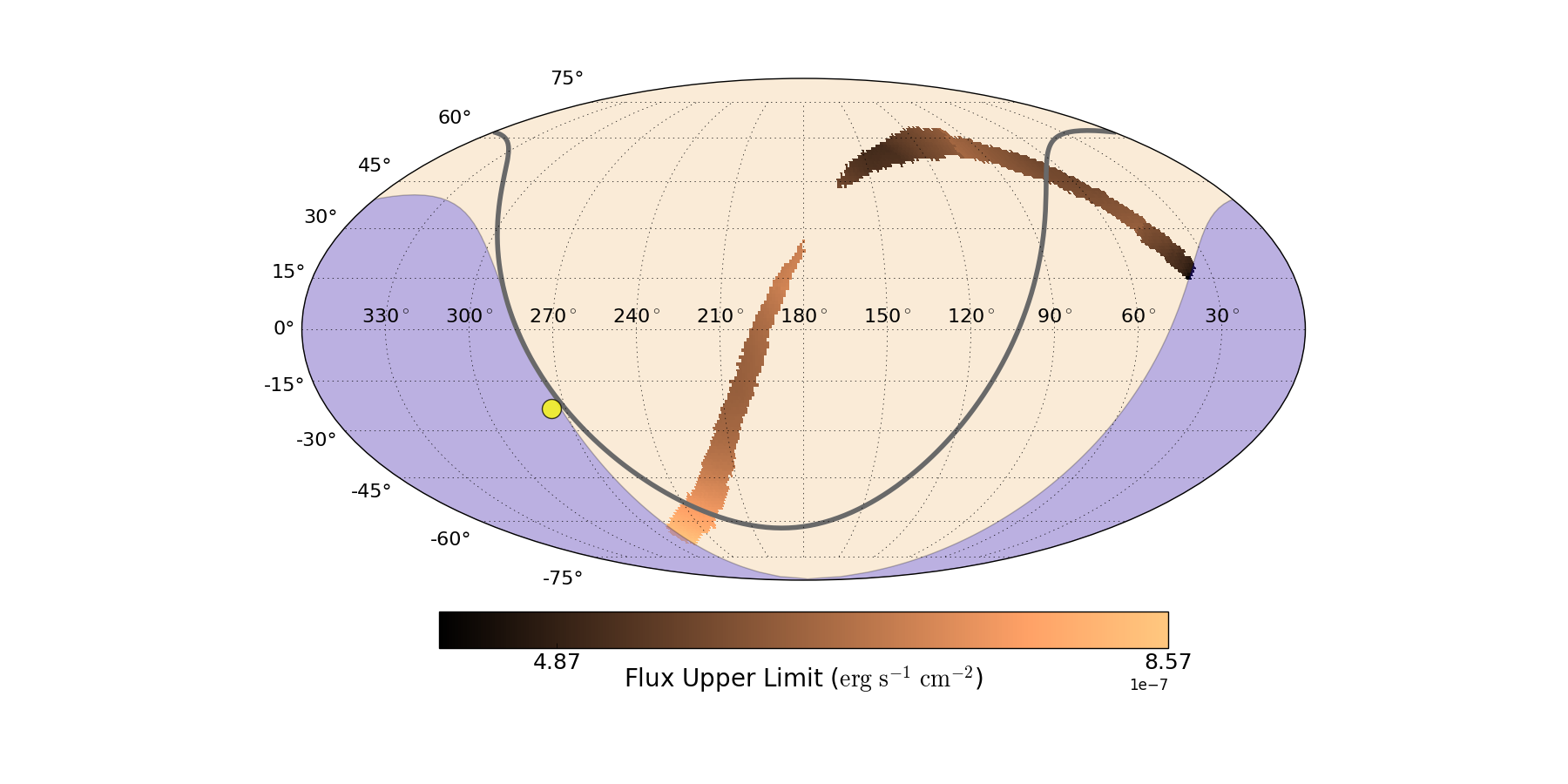}
\caption{The area within the GW151226 LIGO localization contour is shaded to indicate the GBM 10--1000 keV flux upper limits during the the $T_{\rm seeded}$ interval: $-30<t_{\rm GW}<+30$ s.  The purple shaded region indicates where the sky was occulted by the Earth for \Fermi.  The Galactic plane is the grey curve, and the Sun is indicated by the yellow disk.}
\label{fig:GBMGW151226upperlimits}
\end{figure}

We use the same method to calculate the upper limits as for LVT151012. The resulting upper limit map is shown in Figure \ref{fig:GBMGW151226upperlimits}. Using the EOT, we also searched for longer-lasting emission: on timescales of 1 day before and 1 day after $t_{\rm GW}$, 1 month starting at $t_{\rm GW}$ (2015 December 26 to 2016 January 25), and 1 year around $t_{\rm GW}$ (2015 April 28 to 2016 April 28 - shifted to start at $t_{\rm GW}$-242 days and end at $t_{\rm GW}$+124 days - given the data available at the time of this analysis). No new sources were detected on any of the searched timescales and energy bands.

\subsection{LAT}\label{sec:lat_analysis}

The LAT is a pair conversion telescope comprising a $4\times4$ array of silicon strip trackers and cesium iodide (CsI) calorimeters covered by a segmented anti-coincidence detector to reject charged-particle background events. The LAT covers the energy range from 20\,MeV to more than 300\,GeV with a FoV of $\sim 2.4$ sr, observing the entire sky every two orbits ($\sim$3 hours) by rocking north and south about the orbital plane on alternate orbits \citep{2009ApJ...697.1071A}.

sGRBs at LAT energies are often slightly delayed in their onset, have substantially longer durations and appear to come from a different emission component with respect to their keV-MeV signals \citep{latgrbcat1,latgrbcat2}.  The late-time $\gamma$-ray emission has been shown to be consistent with originating from the same emission component as broadband (radio to X-ray) afterglows \citep{grb090510,grb110731,kouveliotou13}. This warrants the search for a high-energy $\gamma$-ray counterpart for GW events on timescales typical of these afterglows (few ks), longer than the prompt emission of an sGRB. Thanks to its survey capabilities, the LAT is well suited to look for such signals.
In addition, given the great uncertainty on the nature of EM signals from BBH mergers, we also search the LAT data over intervals that are much longer than the timescales associated with the afterglow emission of sGRBs, similar to the LAT analysis performed for GW150914 \citep{GW150914_LAT}.

We searched the LAT data for evidence of new transient sources. Since we did not find any evidence of new sources coincident with the LIGO detections, we set flux upper limits (at 95\% c.l.) on the $\gamma$-ray emission in the energy range 100 MeV -- 1 GeV. 

Our analysis is based on the standard unbinned maximum likelihood technique used for LAT data analysis\footnote{\url{http://fermi.gsfc.nasa.gov/ssc/data/analysis/documentation/Cicerone}}.
We include in our \textit{baseline} likelihood model all sources (point-like and extended) from the LAT source catalog (3FGL, \citealt{3fgl}), as well as the Galactic and isotropic diffuse templates provided by the \Fermi-LAT Collaboration \citep{2016ApJS..223...26A}. We employ a likelihood-ratio test \citep{Neyman1928} to quantify whether the existence of a new source is statistically warranted.  In doing so, we form a test statistic (TS) that is two times the logarithm of the ratio of the likelihood evaluated at the best-fit model parameters when including a candidate point source at a given position (alternative hypothesis), to the likelihood evaluated at the best-fit parameters under the baseline model (null hypothesis).
As is standard for LAT analysis, we choose to reject the null hypothesis when the TS is greater than 25, roughly equivalent to a $5 \sigma$ rejection criterion for a single search.

In the following, unless stated otherwise, a point in the sky is considered observable by the LAT if it is within  65$^\circ$ of the LAT boresight (or z-axis) and has an angle with respect to the local zenith smaller than 100$^\circ$. The latter requirement is used to exclude contamination from terrestrial $\gamma$-rays produced by interactions of cosmic rays with the Earth's atmosphere.

We now describe briefly the different searches we have performed. First our \textit{Fixed Time Window search} is used to search for a possible counterpart and provide a global upper limit on the average flux for a fixed time window. This upper limit is relevant if the only information known about the position of a possible counterpart is the LIGO localization map, which is used as a prior. Then, our \textit{Adaptive time binning search} is used to search for counterparts on different time scales, and to provide a map of upper limits which refer to time windows optimized for each location in the map. These limits are useful if, after the publication of this paper, a localization of a potential counterpart more accurate than the LIGO localization map become available. We refer the reader to Vianello et al. (2016, in preparation) for more details about these analyses. 

The results of these analyses for LVT151012 and GW151226 are presented in the following sub-sections.

\subsubsubsection{Fixed Time Window Search}
\label{sec:fixed-time-search}

In this analysis we search for high-energy $\gamma$-ray emission on a set of fixed time windows ($T_{\rm fixed}$), starting at or slightly before the time of the LIGO triggers. 
For each time window, we start by selecting all pixels \citep[the LIGO localization maps are in HEALPix\footnote{\url{http://healpix.sourceforge.net}} format;][]{HEALPix} that were observable by the LAT within the 90\% containment of the LIGO localization maps, down-scaled to a resolution which matches the LAT point-spread function at 100 MeV ($\sim$$4^\circ$; \textit{nside}$= 128$). We then perform an independent likelihood analysis for each pixel, where we test for the presence of a new source at the center of the pixel. For all these likelihood analyses we use the Pass 8 \texttt{P8\_TRANSIENTR010E\_V6} event class and the corresponding instrument response functions\footnote{\url{http://fermi.gsfc.nasa.gov/ssc/data/analysis/documentation/Cicerone/Cicerone_LAT_IRFs/IRF_overview.html}}. Since we did not detect any new source above our TS threshold in any of the positions, we proceeded with the computation of upper limits with the technique detailed in Vianello et al. (2016, in preparation). In short, the LIGO probability map is used as a prior on the position of the EM counterpart, and the posterior probability for its flux $F$ is computed by marginalizing the full posterior with respect to the position and all the other free parameters. We then compute the upper limit for a given probability $p$ as the upper bound of the credibility interval for $F$ which starts at $F=0$ \citep{Agashe:2014kda}.

\subsubsubsection{Adaptive time window search}

In this analysis we optimize the time window for the analysis for each pixel, defining an ``adaptive'' interval ($T_{\rm adaptive}$) that starts when the pixel becomes observable by the LAT (its angle from the LAT boresight is $<$65$^\circ$ and has a Zenith angle $<$92$^\circ$--taking into account the 8$^\circ$ radius Region of Interest, RoI) and ends when it is no longer observable by the LAT.  We further down-scale the HEALPix map (\textit{nside}$= 64$), and for each pixel we select only the interval that contains the GW trigger time, or the one immediately after (if the center of the pixel was outside the LAT FoV at the GW trigger time). This analysis is therefore optimized for the assumption that the source emitted $\gamma$ rays at the time of the GW event, and the time window of the analysis is designed to only contain a continuous observation.
As in the fixed-time window analysis, we perform an independent likelihood analysis for each pixel, where we test for the presence of a new source at the center of the pixel. We use the Pass 8 \texttt{P8\_TRANSIENTR010E\_V6} event class and the corresponding instrument response functions. We found no significant excesses, and therefore compute flux upper limits.

In addition, similar to our analysis for GW150914 \citep{GW150914_LAT}, we search for a significant excess using adaptive time windows but on longer timescales, from 10 days before to 10 days after the GW event.
To limit the number of trials, in this analysis we use a coarser spatial resolution (\textit{nside}$= 8$) that roughly matches the size of each RoI, but we compute the TS map (using the \texttt{ScienceTool}\footnote{\url{http://fermi.gsfc.nasa.gov/ssc/data/analysis/}} \texttt{gttsmap}) for each RoI. As described in \cite{GW150914_LAT}, we use \texttt{gtsrcprob} to assign, to each event, the probability that the event belongs to each of the sources in the likelihood model. We then compute the number of photons that are associated with the source with a probability $>$0.9. This is useful for filtering the excesses caused by random spatial coincidence of single high-energy events from persistent sources or background.

    
\subsubsubsection{Other Standard LAT Searches}

The \Fermi Automatic Science Processing (ASP) pipeline, which is used to search for transient sources (e.g., blazar flares) as regularly reported in Astronomer's Telegrams, was also employed around the GW trigger times.  The ASP pipeline performs a blind search for sources on all-sky count maps constructed from the event data acquired on 6 hr and 24 hr timescales ($T_{\rm ASP}$).  Candidate flaring sources are then fit using a standard likelihood analysis modeled along with known sources and the Galactic and isotropic diffuse contributions.  These candidate sources are then characterized and matched to known sources, allowing for the identification of flaring cataloged sources as well as new unassociated sources.

The \Fermi All-Sky Variability Analysis (FAVA) was also employed to search for excess emission on week-long timescales ($T_{\rm FAVA}$).  The FAVA weeks are pre-defined: therefore we search the week that includes the corresponding LIGO triggers, and the week afterward. FAVA is a blind photometric analysis in which a grid of regions covering the entire sky is searched for deviations from the expected flux based on the observed long-term mission-averaged emission \citep{fava1}.  This allows the FAVA search to be independent of any model of the $\gamma$-ray sky, and therefore complement the standard likelihood analyses. 

\subsubsection{LAT Observations of LVT151012}

\begin{figure}
\centering
\includegraphics[width=0.5\textwidth,trim=20 0 0 0,clip=true]{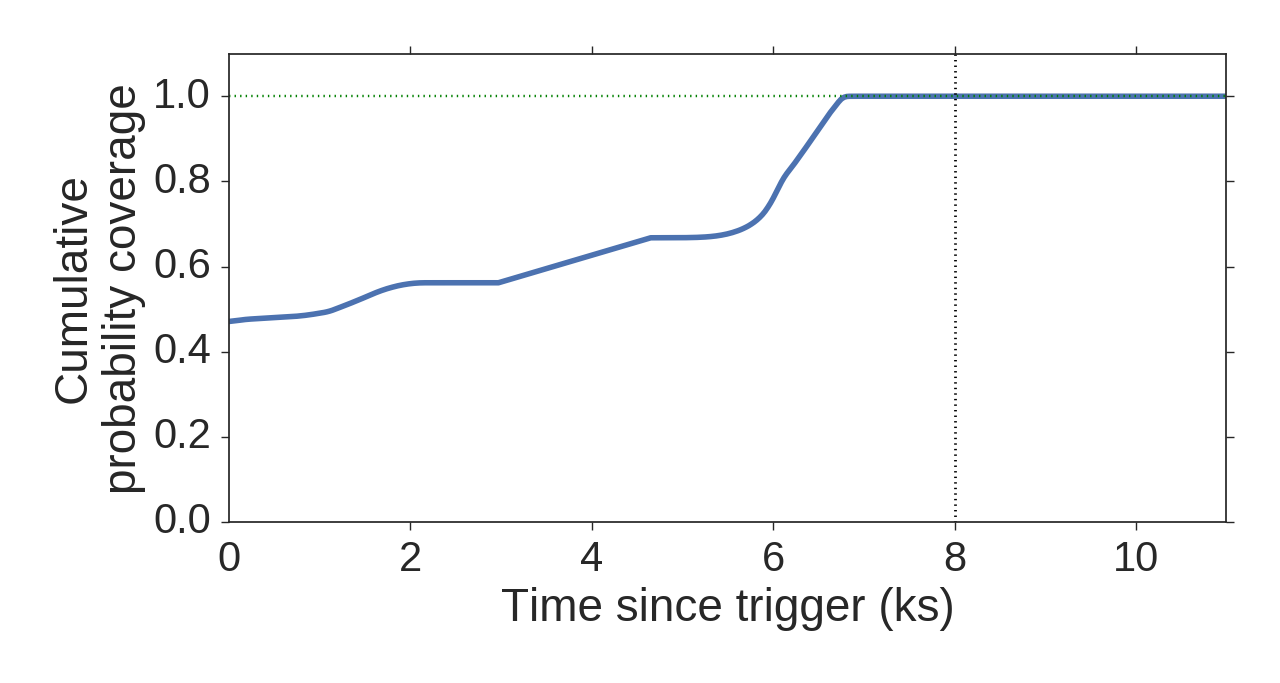}\\
\includegraphics[width=0.5\textwidth,trim=20 0 0 0,clip=true]{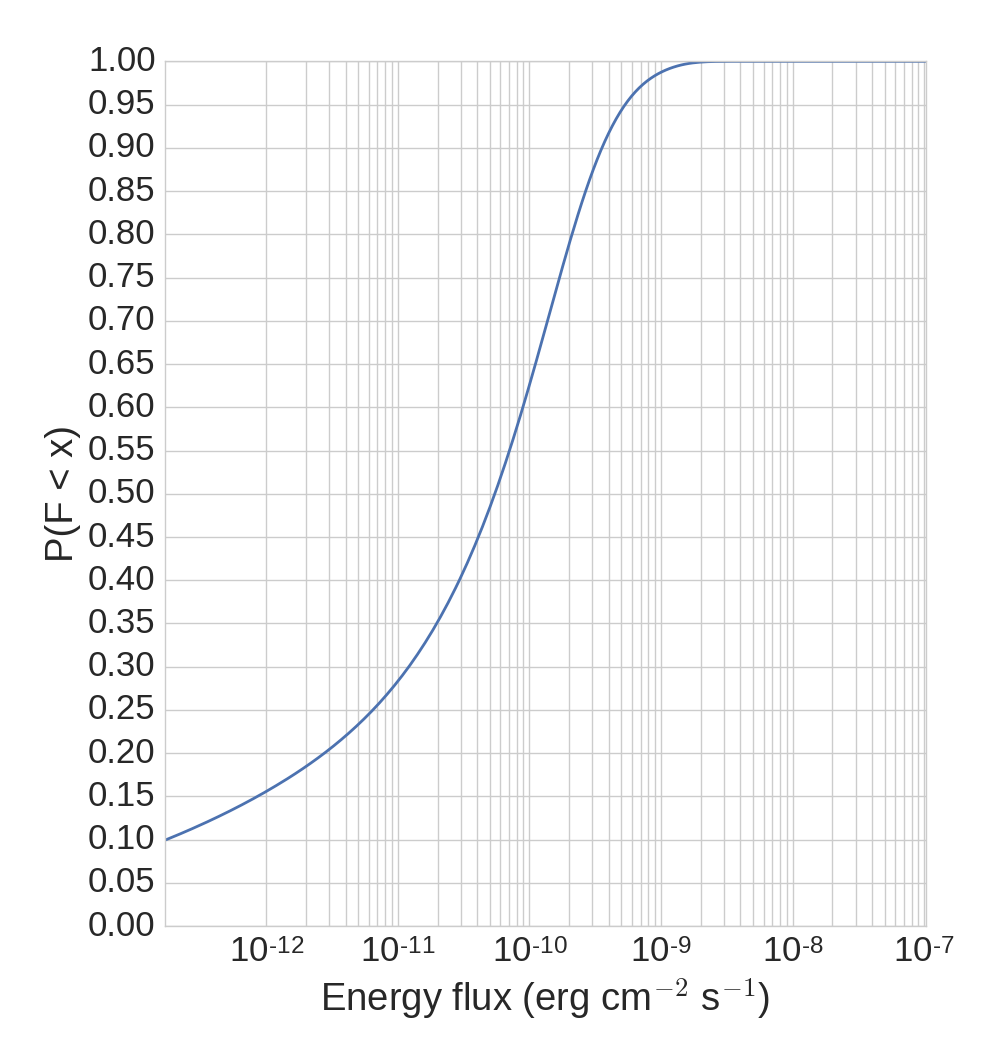}
\caption{The LAT observations of LVT151012: cumulative fraction of the LIGO localization probability observed by the LAT as a function of time since $t_{\rm LVT}$ ({\it top}); Integral of the marginalized posterior for the 0.1--1 GeV energy flux ({\it bottom}) during the $T_{\rm fixed_2}$ interval. The flux at which the blue curve intersects a given probability $P(F < x)$ corresponds to the upper limit at that credibility level.}
\label{fig:lat_lvt151012}
\end{figure}

\Fermi was in sky survey mode at the time of the GW signal from LVT151012, $t_{\rm LVT}$, rocked 50$^{\circ}$ South from the orbital plane. The LAT was favorably oriented toward LVT151012, covering $\sim$47\% of the LIGO localization probability at the time of the trigger.  Within $\sim 7$~ks from $t_{\rm LVT}$, the LAT had observed $\sim$100\% of the LIGO localization probability (Figure \ref{fig:lat_lvt151012}).  
The LAT then continued to observe the entire LIGO localization region throughout normal sky-survey operations in the days and months afterward.

We performed the fixed time window search described in Section \ref{sec:fixed-time-search} on two time intervals. The first interval $T_{\rm fixed_1}$ covered from $t_{\rm LVT} - 10 s$ to $t_{\rm LVT} + 10$ s, during which the LAT observed $\sim$50\% of the LIGO localization map, corresponding to almost the entire Southern region of the LIGO localization contour. This search is relevant for finding high-energy emission close in time and of similar duration with respect to the GW signal. The second time window $T_{\rm fixed_2}$ covered from $t_{\rm LVT}$ to $t_{\rm LVT} + 8$~ks, which corresponds to the time interval when the LAT fully observed the LIGO localization map (see Figure \ref{fig:lat_lvt151012}), plus $1$ ks to accrue some exposure of the final regions that became visible to the LAT. We found no credible candidate counterparts in $T_{\rm fixed_1}$ or in $T_{\rm fixed_2}$. We then performed the upper limit computation described in Section \ref{sec:fixed-time-search} for $T_{\rm fixed_2}$. The integral function of the marginalized posterior for the 0.1--1 GeV energy flux is shown in the bottom panel of Figure~\ref{fig:lat_lvt151012}. This function can be used to evaluate the upper limit for different credibility levels. In particular, the 95\% upper limit is $F_{ul, 95} = 5 \times 10^{-10}$ erg cm$^{-2}$ s$^{-1}$. 

\begin{figure*}
\centering
\includegraphics[width=0.65\textwidth,trim=100 0 100 25,clip=true]{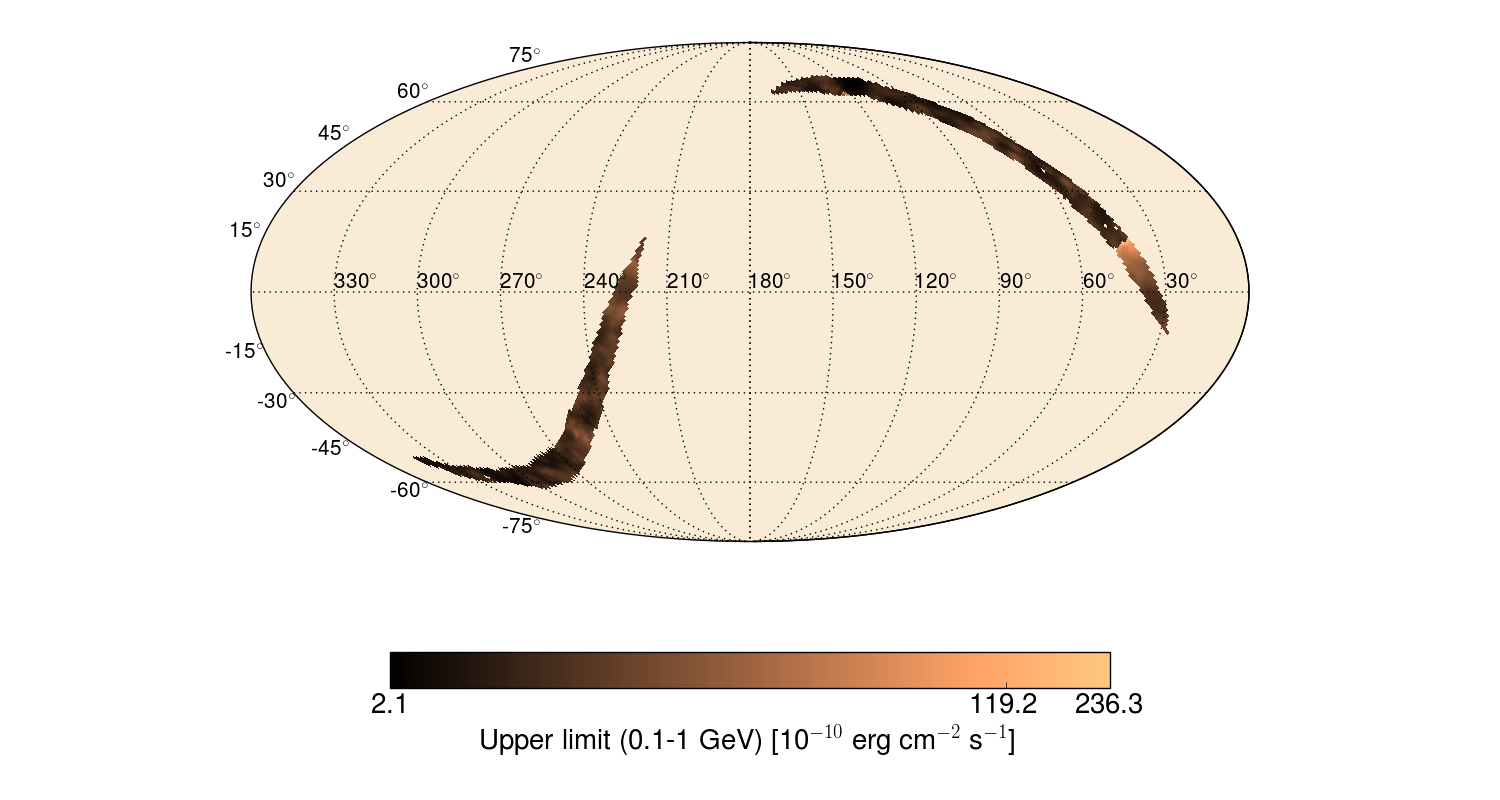}
\includegraphics[width=0.55\textwidth,trim=0 30 30 10,clip=true]{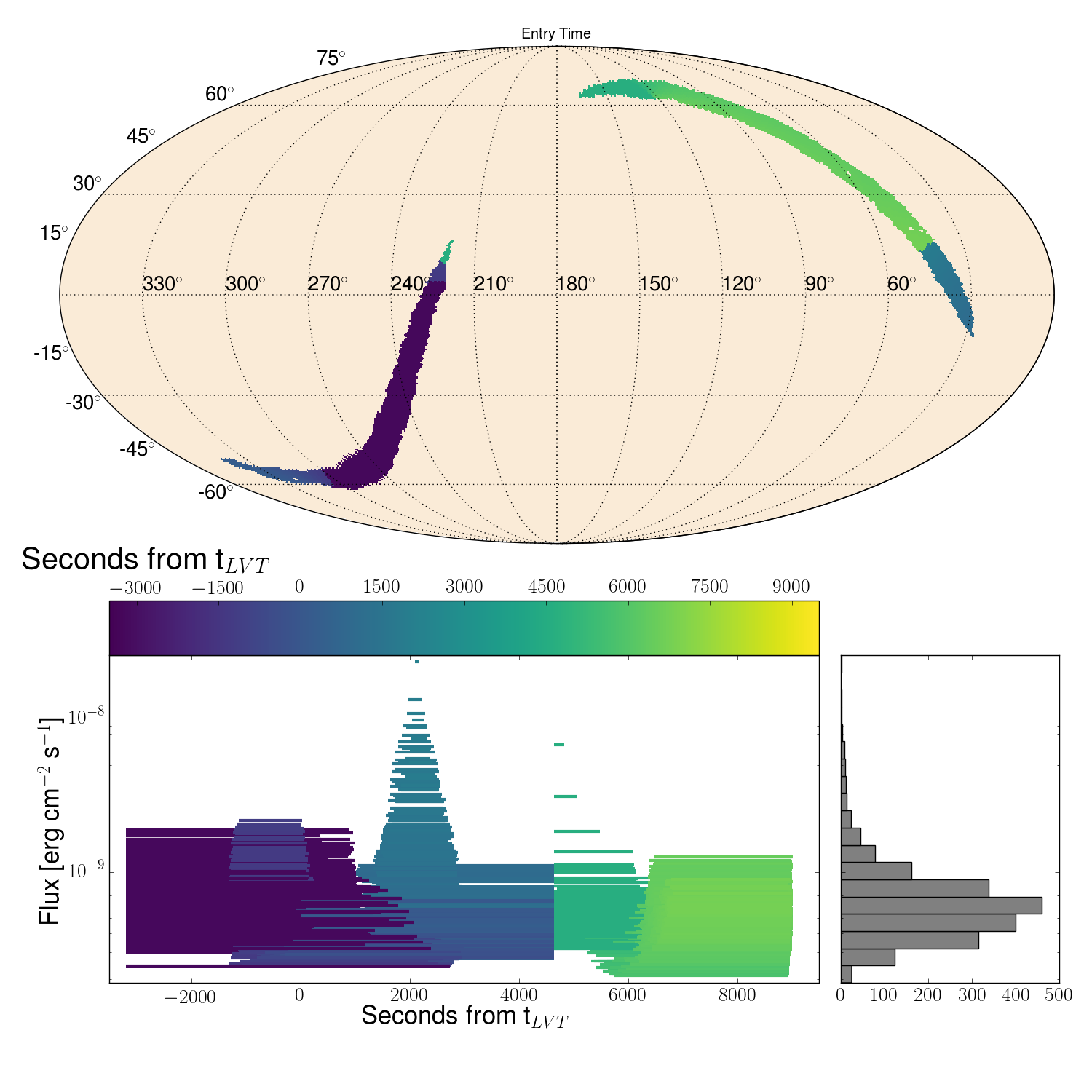}
\caption{The adaptive time interval analysis for LVT151012 over the first \Fermi orbit containing $t_{\rm LVT}$: Flux upper limit map during $T_{\rm adaptive}$ ({\it top}), the entry time into the LAT FoV relative to $t_{\rm LVT}$ of the RoI for each pixel within the LIGO localization contour ({\it middle}), and the upper limit light curves for each RoI ({\it bottom}).
The horizontal bars in the bottom panel correspond to the values of the LAT upper limits, and their position along the time-axis coincides with the interval of time used in the analysis. 
The color of each bar indicates the time when the RoI entered the LAT FoV, and matches  the color of the pixel in the middle panel. The horizontal histogram displays the distribution of upper limits.}
\label{fig:lat_ul_151012_adap}
\end{figure*}

The adaptive time window ($T_{\rm adaptive}$) analysis did not yield any significant excesses, and no new sources were detected above a likelihood detection threshold of TS=25, neither in the time window containing or just after $t_{\rm LVT}$, nor in a scan of 10 days before and 10 days after $t_{\rm LVT}$. The flux upper limits for the portion of the LIGO localization contour containing 90\% of the probability during the adaptive time window are shown in Figure \ref{fig:lat_ul_151012_adap}. The values for the flux upper limits from this analysis range from 2.1 $\times 10^{-10}$ erg~cm$^{-2}$~s$^{-1}$ up to 2.4 $\times 10^{-8}$ erg~cm$^{-2}$~s$^{-1}$. Most of the flux upper limits are below $10^{-9}$ erg~cm$^{-2}$~s$^{-1}$, and the tail extending to higher fluxes is due to a region with poor exposure of the LIGO contour (at RA$\simeq 30 ^\circ$, Dec$\simeq0 ^\circ$) entering the LAT FoV at approximately $t_{\rm LVT}+2$ ks.


We examined the ASP products during the 6 hour and 24 hour intervals ($T_{\rm ASP}$) containing $t_{\rm LVT}$.  No new unassociated flaring sources were detected within the LIGO localization contour.

The FAVA search encompassed a pre-defined week before and after $t_{\rm LVT}$ ($T_{\rm FAVA}$); the weeks from 2015 October 5 to 2015 October 12, and 2015 October 12 to 2015 October 19, respectively.  The FAVA search over these two periods detected a total of five flaring sources above 5$\sigma$ within the LIGO localization region.  A followup likelihood analysis of each seed flare determined their positions to be consistent with known flaring 3FGL sources.



\subsubsection{LAT Observations of GW151226}

\begin{figure}
\centering
\includegraphics[width=0.5\textwidth,trim=20 0 0 0,clip=true]{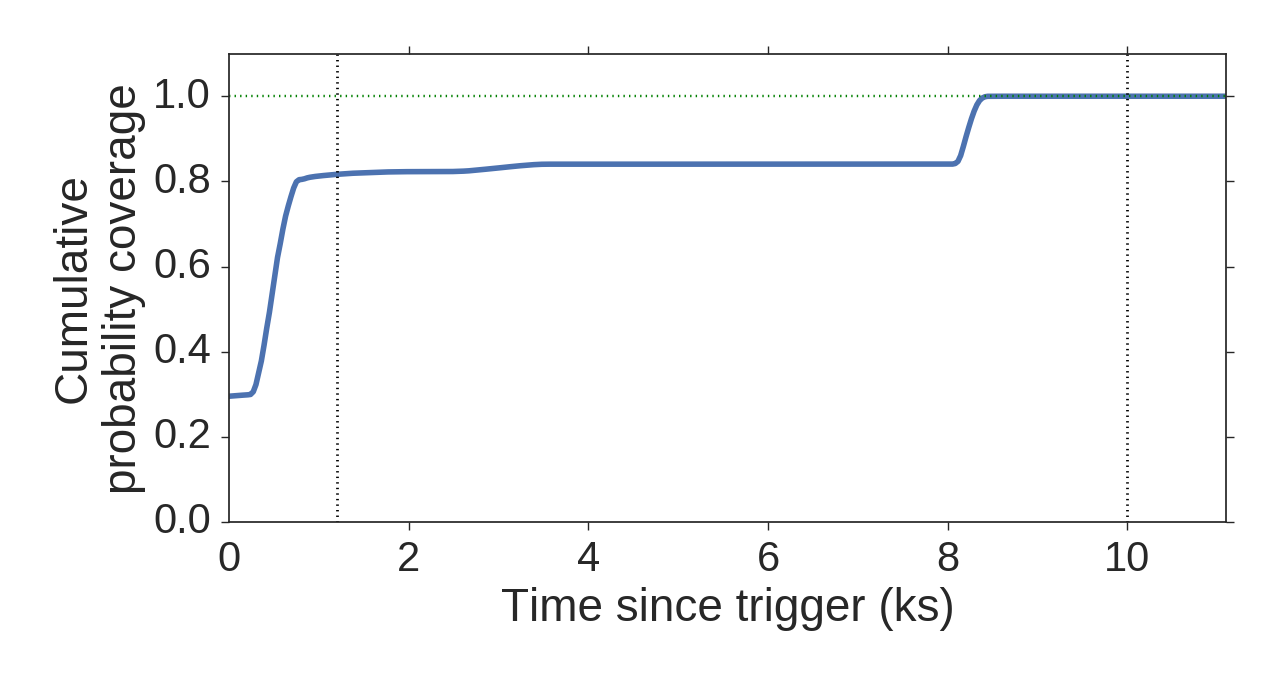}\\
\includegraphics[width=0.5\textwidth]{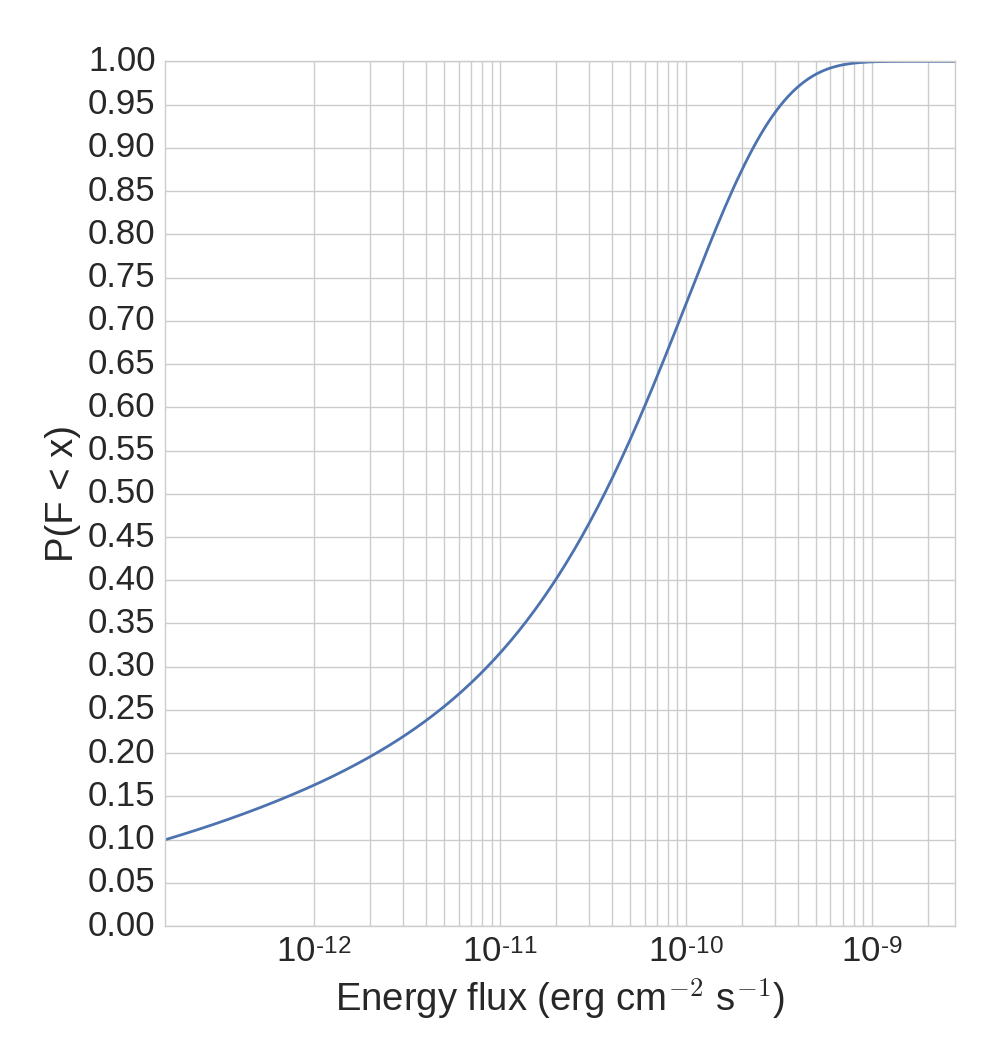}\\
\caption{The LAT observations of GW151226: cumulative fraction of the LIGO localization probability observed by the LAT as a function of time since $t_{\rm GW}$ ({\it top}); Integral of the marginalized posterior for the 0.1--1 GeV energy flux ({\it bottom}) during the $T_{\rm fixed_3}$ interval. The flux at which the blue curve intersects a given probability $P(F < x)$ corresponds to the upper limit at that credibility level.}
\label{fig:lat_gw151226}
\end{figure}

\Fermi was in sky survey mode at the time of the GW detection of GW151226 (03:38:53.648 UTC on 2015 December 26, $t_{\rm GW}$ in the following), rocked 50$^{\circ}$ North from the orbital plane.
The LAT was favorably oriented toward GW151226, covering $\sim$32\% of the LIGO localization probability at the time of the trigger.  Within $\sim 1$~ks from $t_{\rm GW}$ the LAT had observed $\sim$80\% of the LIGO localization probability, and $\sim$100\% within $\sim t_{\rm GW}$+8.5 ks (Figure \ref{fig:lat_gw151226}). The LAT continued to observe the entire LIGO localization region throughout sky-survey operations in the days and months afterwards.  

We performed the fixed time window search described in Section~\ref{sec:fixed-time-search} on three time intervals. The first interval $T_{\rm fixed_1}$ covered from $t_{\rm GW} - 10 s$ to $t_{\rm GW} + 10$ s. During $T_{\rm fixed_1}$ the LAT observed $\sim$30\% of the LIGO localization probability. The second interval $T_{\rm fixed_2}$ covered from $t_{\rm GW}$ to $t_{\rm GW} + 1.2$~ks, which corresponds to a shorter time interval when the LAT had an appreciable fractional coverage ($\sim 80$\%) of the LIGO localization probability (see Figure \ref{fig:lat_gw151226}), with $200$ s added to accrue some exposure at the final regions to become visible to LAT. The third interval $T_{\rm fixed_3}$ covered from $t_{\rm GW}$ to $t_{\rm GW} + 10$~ks, and corresponds to the time interval during which the LAT had 100\% coverage, with $\sim 1$~ks added to accrue some exposure for the final points to become visible to the LAT. We found no candidate counterpart in any of the three time windows. We then performed the upper limit computation described in Section \ref{sec:fixed-time-search} for $T_{\rm fixed_3}$. The integral function of the marginalized posterior for the 0.1--1 GeV energy flux is shown in the bottom panel of Figure~\ref{fig:lat_gw151226}. The 95\% upper limit is $F_{ul, 95} = 3 \times 10^{-10}$ erg cm$^{-2}$ s$^{-1}$. 

\begin{figure*}
\centering
\includegraphics[width=0.65\textwidth,trim=100 0 100 25,clip=true]{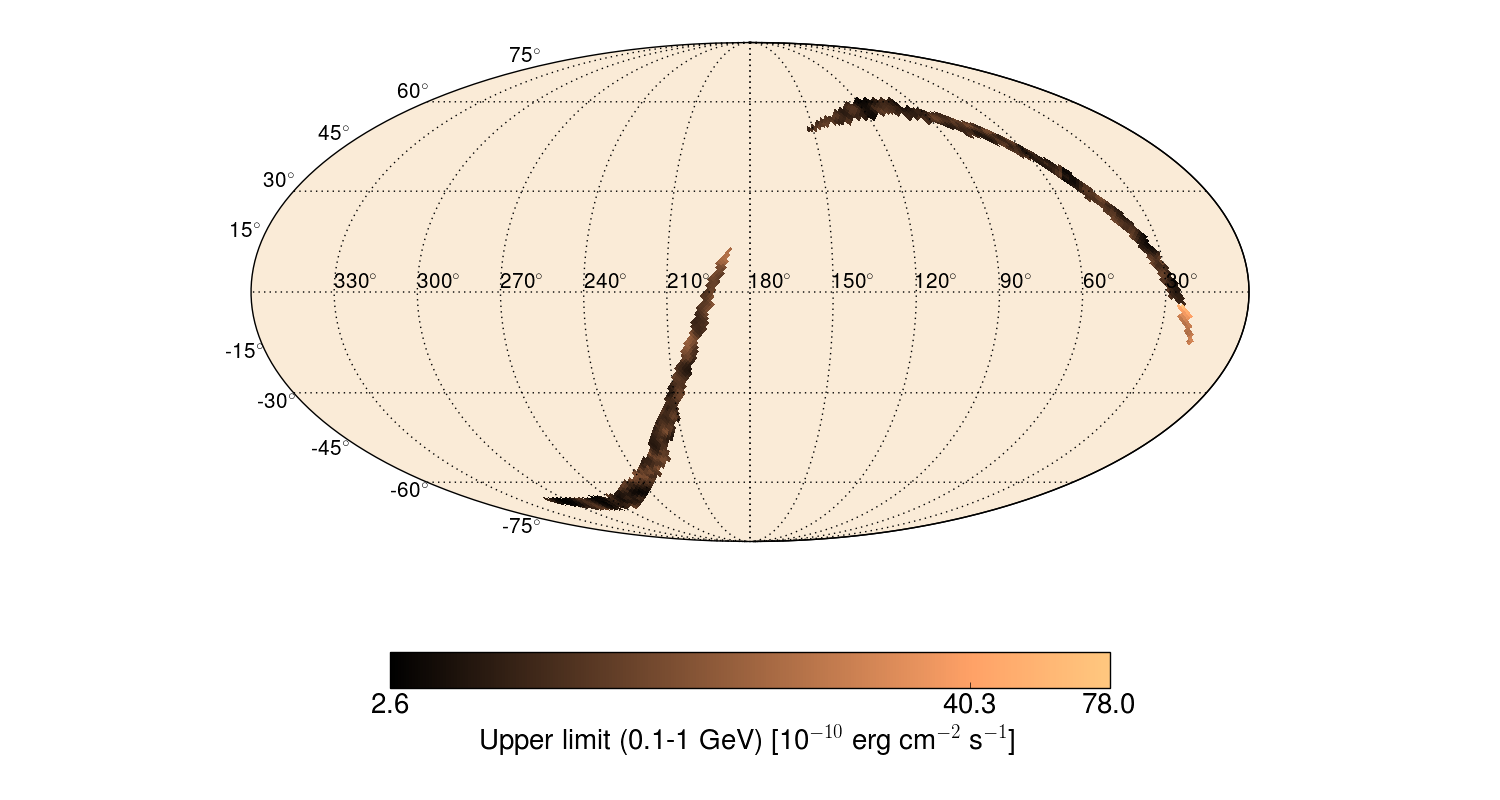}
\includegraphics[width=0.55\textwidth,trim=0 30 30 10,clip=true]{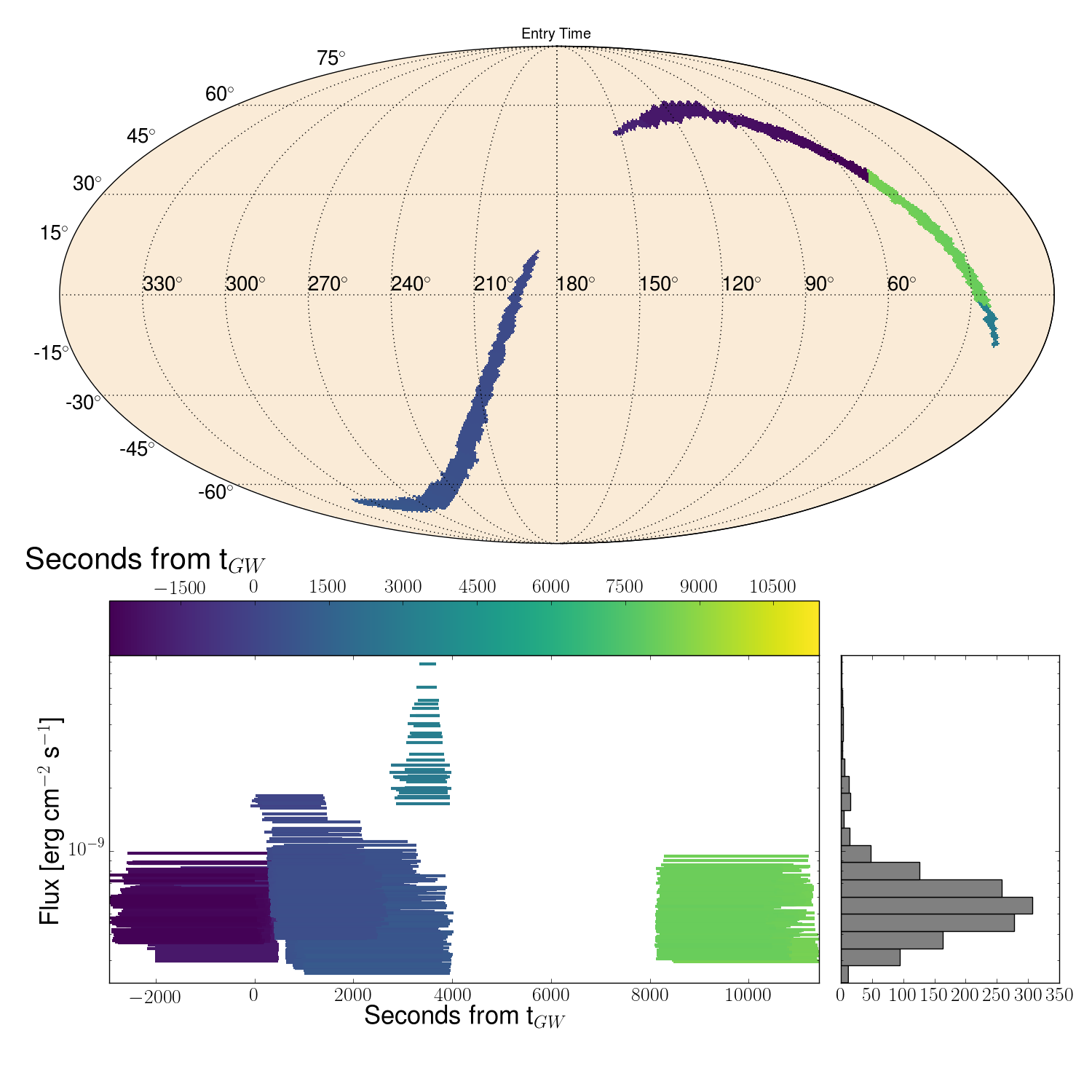}
\caption{The adaptive time interval analysis for GW151226 over the first \Fermi orbit containing $t_{\rm GW}$: Flux upper limit map during $T_{\rm adaptive}$ ({\it top}), the entry time into the LAT FoV relative to $t_{\rm GW}$ of the RoI for each pixel within the LIGO localization contour ({\it middle}), and the upper limit light curves for each RoI ({\it bottom}).
The horizontal bars in the bottom panel correspond to the values of the LAT upper limits, and their position along the time-axis coincides with the interval of time used in the analysis. The color of each bar indicates the time when the RoI entered the LAT FoV, and matches the color of the pixel in the middle panel. The horizontal histogram displays the density of upper limits.}
\label{fig:lat_ul_151226_adap}
\end{figure*}

The adaptive time window analysis did not lead to the detection of any new $\gamma$-ray sources during the first \Fermi orbit ($\sim$96 minutes) after $t_{\rm GW}$. 
The results of this analysis are shown in Figure \ref{fig:lat_ul_151226_adap}. 
For this event, the values for the flux upper limits range from 2.6 $\times 10^{-10}$ erg~cm$^{-2}$~s$^{-1}$ up to 7.8 $\times 10^{-9}$ erg~cm$^{-2}$~s$^{-1}$. 
The tail of upper limits with values $>10^{-9}$ erg~cm$^{-2}$~s$^{-1}$ is due to a series of regions entering the FoV at about 3 ks (at R.A.$\simeq30^\circ$, Dec$\simeq-15^\circ$) for which the exposure was particularly low. 
As for the previous event, we also searched for excess $\gamma$-ray emission on an orbit-by-orbit timescale over $\pm$10 days on either side of $t_{\rm GW}$. 
No new sources were detected above a threshold of TS=25.  The most significant flaring source within the LIGO localization region is the blazar PKS 1424-41, which has flared regularly over the entire \Fermi mission\footnote{\url{http://fermi.gsfc.nasa.gov/ssc/data/access/lat/FAVA/SourceReport.php?week=386&flare=31}}$^,$\footnote{\url{http://fermi.gsfc.nasa.gov/ssc/data/access/lat/msl_lc/source/PKS_1424-41}}.




We examined the ASP products during the 6 hour and 24 hour intervals containing $t_{\rm GW}$.  No new unassociated flaring sources were detected within the LIGO localization contour.

The FAVA search for emission associated with GW151226 encompassed the pre-defined weeks of 2015 December 21 to 2015 December 28 and 2015 December 28 to 2016 January 4, and detected a total of five flaring sources above 5$\sigma$ within the LIGO localization region. Again, a dedicated followup likelihood analysis of each seed flare determined their positions to be consistent with known flaring 3FGL sources, including the highly active blazar PKS 1424-41 (3FGL J1427.9-4206).




\section{Discussion}\label{sec:disc}
\subsection{Implications for Candidate Counterpart GW150914-GBM}
The candidate $\gamma$-ray counterpart to GW150914 reported by the GBM \citep{GW150914_GBM} that resembles a weak sGRB has surprised the community and also spurred a great deal of theoretical speculations.  The low significance of the signal, and the lack of corroboration by other experiments has caused the true nature of the GBM signal to remain ambiguous (see also \citealt{greiner16,GW150914_integral}). Strong support for the candidate EM counterpart would be achieved if a similar or higher significance counterpart were found associated with other GW BBH merger events.  
The \Fermi non-detections of $\gamma$-ray counterparts to LVT151012 and GW151226 can neither confirm nor refute the potential association between GW150914 and the GBM candidate counterpart.

If we assume that all BBH mergers produce sGRB-like signals, the GBM might reasonably not detect them for four reasons:
\begin{itemize}
\item The GBM observed only 68\% and 83\% of the LIGO localization probability of LVT151012 and GW151226, respectively, at the times of the GW triggers.  Therefore, there is a significant probability that the LIGO sources could have been simply occulted by the Earth for \Fermi at the GW trigger times. Without all-sky coverage by the detecting instrument or a set of identical detectors, a non-detection cannot rule out this hypothesis without a sample much larger than the three events from the LIGO O1 observing run. The fractional sky coverage alone can account for having a single detection.
\item Depending on the source location, orientation, and geomagnetic coordinates of \Fermi at the time of the GW trigger, the GBM background rates can vary substantially. The background count rates were a few hundred Hz higher (3\%) at the time around GW151226 and a few thousand Hz higher (18\%) at the time around LVT151012 than around the time of GW150914. The reported distance to LVT151012 from GW parameters is a factor of $\sim3$ larger than the distance to GW150914. If all of these events produced similar $\gamma$-ray luminosities, the counterpart to LVT151012 would have been indistinguishable from background. 
\item If the source producing $\gamma$ rays in GW150914 is collimated, only a fraction of those objects would be pointed at the Earth.  This fraction is slightly enhanced by the fact that GW signals from binary mergers, while not truly collimated, have stronger GW emission along the rotation axis of the merger system, which is presumably aligned with the EM jet collimation axis.  If one assumes that BBH merger counterparts are collimated similarly to sGRBs \citep{fong15}, then only $\sim$15--30\% 
of similar systems would have their $\gamma$-ray jets pointed toward Earth.  The potential detection of a counterpart in one of three objects is entirely consistent with the most conservative assumptions of the degree to which the high-energy emission from such sources is collimated.
\item The intrinsic luminosity distribution also limits detectability. Even if GW151226 was beamed and on-axis, and the progenitor was not occulted to \Fermi, the event still may not be detectable if it was intrinsically dimmer than GW150914-GBM. The energy radiated as GWs scales strongly with total progenitor mass. If there is also a strong scaling between total progenitor mass and the energy radiated in $\gamma$ rays, then any $\gamma$-ray emission from GW151226 would likely be less luminous than that from GW150914.
\end{itemize}

With only three possible GW detections (one with a fairly high FAR), and one candidate counterpart, the statistics are not large, and little can be said of these objects other than that they are broadly consistent.  As Virgo joins LIGO in upcoming GW observing runs, and they both head toward design sensitivity, the localization regions are anticipated to become smaller and the GW horizon distance increase (with the rate increasing as a cubed factor, \citealt{2016LRR....19....1A}).   \Fermi will continue to monitor the sky for potential coincident $\gamma$-ray counterparts to all GW source types.


\subsection{Comparison to the sGRBs}


Although the potential for EM counterparts to BBH mergers has not been well established in the literature, recent development spurred by the GBM report of a candidate counterpart to GW150914 (e.g., \citealt{Fraschetti2016}, \citealt{Janiuk2016}, \citealt{Loeb2016}) suggests that mechanisms may exist.  The connection between BNS (or NS-BH) mergers and sGRBs is much stronger than that of BBH mergers \citep{2012ApJ...746...48M, 2013ApJ...767..124N}, supported by extensive observational evidence (host galaxy observations and offsets, environmental densities inferred from GRB afterglow modeling, observational rates; \citealt{2008MNRAS.385L..10T,fong15}), and consistency with numerical modeling (jet production, magnetic fields; \citealt{2005ApJ...634.1202R}, \citealt{Rezzolla2011}).

We do not suggest that GW150914, LVT151012, or GW151226 necessarily produced EM counterparts similar to the population of hundreds of sGRBs observed by BATSE, Konus-Wind, {\it Swift}-BAT, GBM and other instruments over the last five decades.  However, we put our observations (candidate counterpart and upper limits) of these GW detections in the context of the more familiar sGRBs to demonstrate the capability of both the GBM and LAT for these searches in the future.

\begin{figure}[t!]
\centering
\includegraphics[width=0.5\textwidth]{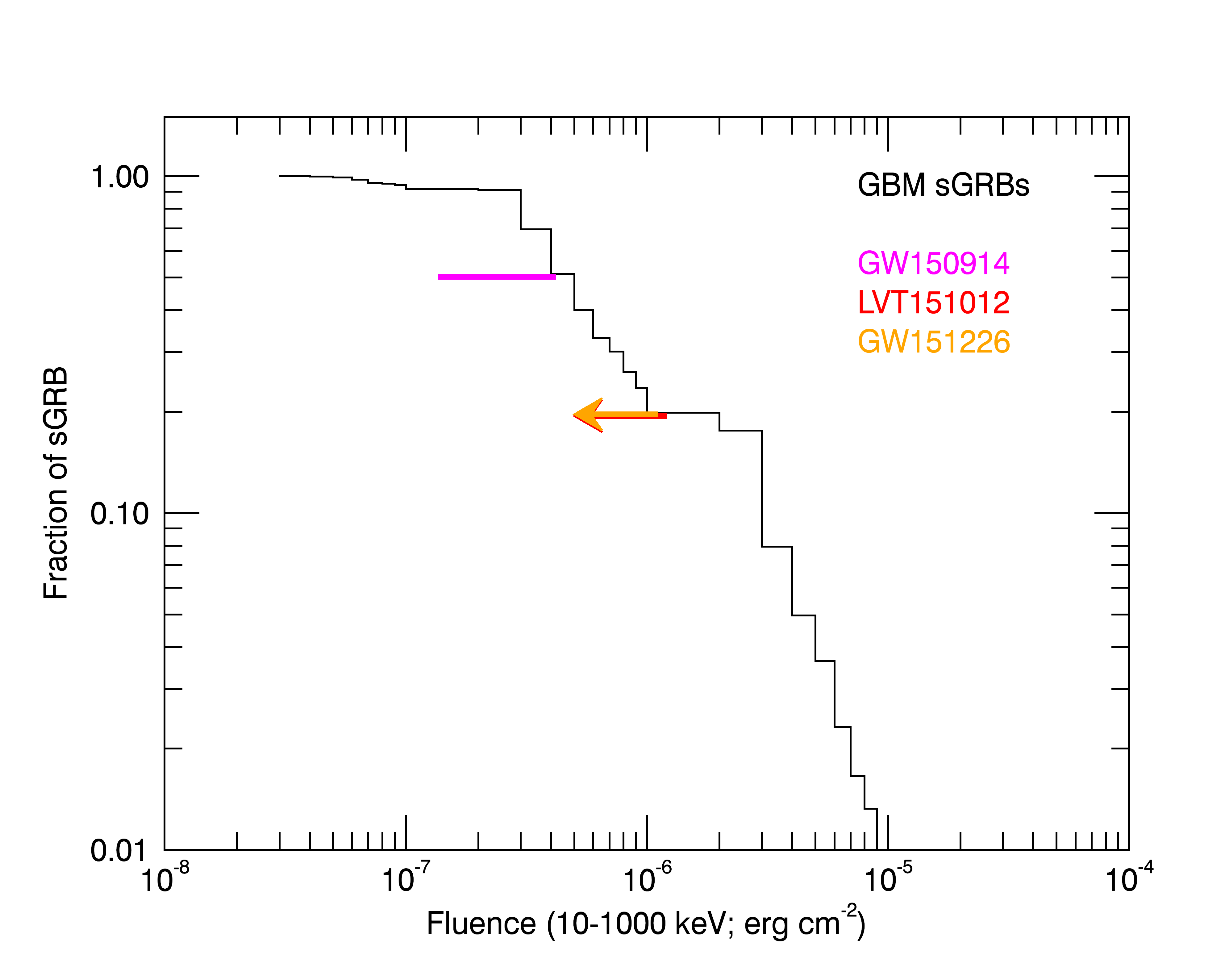}
\caption{Integral distribution of GBM fluence of sGRBs from \cite{3rdgbm} over the duration of the sGRBs, compared to the 1-s fluence measurement for GW150914-GBM, and the upper limits on LVT151012 and GW151226.}
\label{fig:gbm_sgrbs}
\end{figure}

In Figure \ref{fig:gbm_sgrbs}, we compare the distribution of sGRB 1-s fluence measurements from the 3rd GBM GRB catalog \citep{3rdgbm} to our upper limits for LVT151012 and GW151226, as well as the fluence measurement described in \cite{GW150914_GBM}. 
The fluences from the GBM-detected sGRBs span $2.5\times 10^{-8}$ to $1.1\times  10^{-5}$ erg cm$^{-2}$, with GW150914-GBM around the 40th percentile. Compared to sGRBs with known redshifts, GW150914-GBM was unusually close and thus would be very sub-luminous compared to the sGRB population. At a more typical sGRB redshift of $z \sim 0.5$ \citep{2014MNRAS.442.2342D,fong15}, GW150914-GBM would be undetectable by the GBM.  The GBM blind-search reveals sGRB candidates that are a factor of two or three weaker than those triggering the GBM on board.  This opens the possibility of detecting additional fainter sGRBs, and thus testing for the presence of a sub-luminous population that might be associated with BBH mergers (\Fermi-GBM Collaboration 2016, in preparation).

\begin{figure}[t!]
\centering
\includegraphics[width=0.5\textwidth]{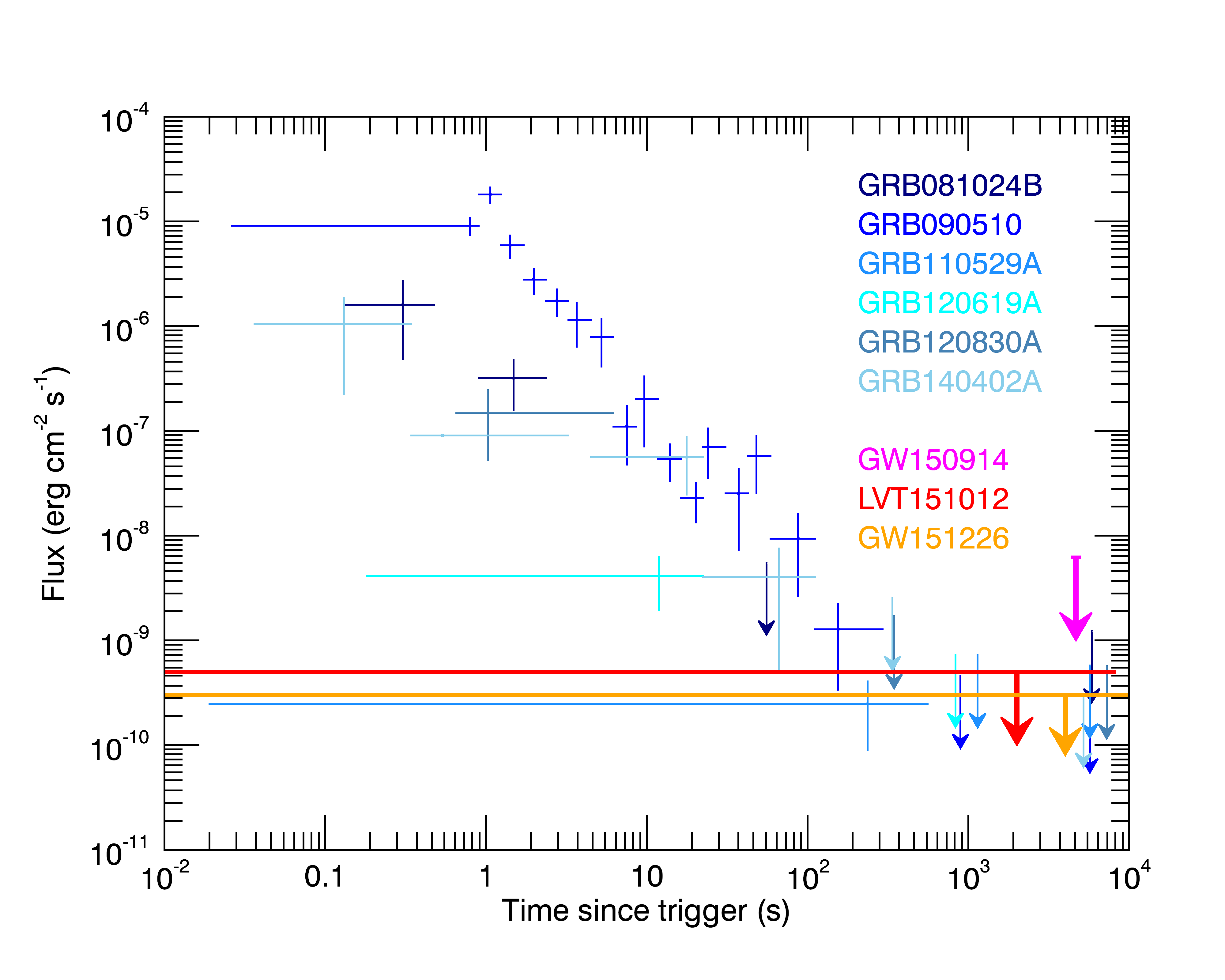}
\caption{A comparison between a selection of the longer-lasting LAT-detected sGRBs with the upper limits from the fixed time-intervals for the three GW events.  The arrows represent the 95\% confidence upper limits from the fixed time windows ($T_1$ from \citealt{GW150914_LAT} for GW150914, $T_2$ for LVT151012, and $T_3$ for GW151226).
}\label{fig:lat_sgrbs}
\end{figure}

The LAT has detected far fewer sGRBs than the GBM, only ten to date in the 100 MeV to $>$300 GeV band, and only the very luminous GRB 090510 \citep{grb090510_lat} has a measured redshift.  In \cite{GW150914_LAT}, we compared the $>$100 MeV light curve of GRB 090510 scaled to the distance inferred from the GW measurements for GW150914 to demonstrate the constraining power of the LAT limits.  We expand that comparison in Figure \ref{fig:lat_sgrbs} to include additional sGRBs (\Fermi-LAT Collaboration 2016, in preparation), and demonstrate that the LAT upper limits from all three GW events are comparable to the measured emission from the LAT-detected sGRBs.  Therefore, if the GW events had extended high-energy $\gamma$-ray emission similar to these sGRBs, it would have been detectable by the LAT within tens-to-hundreds of seconds after the trigger. 

\subsection{Theoretical Insights Concerning EM Counterparts for BBH Mergers}

The excitement of the watershed LIGO discovery has precipitated numerous merger models with EM emission components, ranging from sGRBs to optical and radio transients (e.g., \citealt{Murase2016}) and even luminous neutrino sources (e.g., \citealt{Janiuk2016,2016arXiv160208436M}). This discussion is restricted to an incomplete selection of counterpart models, with a view to defining key observational elements that modelers should address in future studies.

Much of the flurry of very recent activity in GW+EM merger modeling has centered on systems with circumbinary disks or common envelopes that can seed ephemeral accretion onto the resultant BH, perhaps spawning sGRBs. The study of \cite{2016arXiv160300511W} explores the evolution of close binaries composed of massive stars, with core collapse in sequence: one companion generates a BH, and the second one facilitates faster precursor inspiral due to the presence of a common envelope.  After the second BH is formed, the merger takes place amid the ambient shroud that provides fodder for EM emission. Such a picture is adopted by \cite{Janiuk2016} as a basis for their neutrino flux predictions. A different scenario is that of \cite{Loeb2016}, who discusses a single star progenitor for a BBH merger: the rapid rotation of the massive star yields either a dual helium core or ``dumbbell'' core configuration that spawns transient BHs that then merge.  The common envelope again naturally feeds the ergosphere with material for processing into EM form. The model of \cite{2016ApJ...821L..18P} employs an extant BBH system that possesses a residual disk at large radii that is neutral and therefore suppresses the magneto-rotational instability. This ``fallback'' disk remains inert until BBH inspiral revives it through tidal disruption and associated heating. The merger then drives belated accretion to generate an sGRB in temporal connection with the GW event. Even though the focus in these pictures is on the accretion, there is the suggestion that jet activity will be part of the rapidly evolving system.  Winds may also be present (e.g., \citealt{Murase2016}), and the lesser collimation of these can enhance the detectability of energetic EM signals.

A number of the counterpart models invoke the extraction of energy and angular momentum from the ergospheres of the merging BHs via the Blandford-Znajek mechanism \citep{1977MNRAS.179..433B}, a process that is posited to supply matter and energy to the bases of jets emanating from supermassive BHs.  Exploring this possibility in detail is beyond the scope of the present suite of incipient models of mergers.
Yet it should be noted that \cite{Lyutikov2016} and \cite{Murase2016} indicate that the EM luminosity constraints from such EM induction physics for GW150914 may require TeraGauss magnetic fields, with \cite{Lyutikov2016} suggesting that these could be unrealistically large for BH environs.

The challenge for future theoretical studies of BBH mergers generating EM counterparts is to establish EM templates for observational predictions at a fairly detailed level.  These must address typical values and ranges for the source luminosity, multi-wavelength spectrum and angular collimation. They should also offer clear assessments of the pertinent timescales for the events, including delay relative to the GW event and duration in different wavebands, and also whether or not there is EM precursor activity \citep{Ciolfi2015}.  There is also the necessity of establishing a GW merger signal with frequency and frequency derivative character appropriate to the waveforms observed by LIGO and Virgo, i.e. matching oscillatory temporal templates calculated assuming a pair of BHs merging in vacuum. This array of model discriminants will enable rapid progress should GW+EM mergers become an established astronomical paradigm.

Depending on the quality of the counterpart data, it may also be
possible to constrain elements of fundamental physics.  Most notable among a plethora of ideas in the literature is using the time separation between the GW and EM signals to limit departures of the light signal speed from $c$ (e.g., \citealt{2016arXiv160408530B,2016arXiv160308955Y}).  This can potentially constrain Lorentz invariance violations that can be attributed to various physics concepts such as quantum gravity.  Such an enterprise would require significant photon counting statistics and achromatic light curves, as was the case for the GBM+LAT data for the bright burst GRB090510 \citep{grb090510_nature, Vasileiou2013}. The prospect for probing fundamental physics emphasizes the importance of having $\gamma$-ray
monitoring capability in place during the era of advanced GW detectors.

\section{Conclusions}\label{sec:conc}
\Fermi GBM and LAT provide the best current wide-field observations of the time-variable $\gamma$-ray sky in the keV--GeV band, for comparison to triggers from multi-messenger facilities like LIGO.  The GBM and LAT observed a substantial fraction of the LIGO localization probabilities at the times of the LIGO triggers for the three potential BBH mergers, and fully observed them within minutes to hours later. The GBM candidate counterpart for GW150914 and the non-detections from LVT151012 and GW151226, as well as the LAT non-detections for all three merger candidates, can provide observational constraints for new theoretical models for EM counterparts to BBH mergers.

Unfortunately, \Fermi observations of LVT151012 and GW151226 cannot conclusively resolve the unknown nature of the GBM candidate counterpart to GW150914.  The partial GBM and LAT coverage of the LIGO localization regions at the time of trigger for both LVT151012 and GW151226 leaves open the possibility that similar EM counterparts occurred outside the GBM and LAT FoVs.  Ultimately, a statistically large sample of well-observed localization probability maps for BBH mergers will be needed to confidently say whether GW150914-GBM is associated with a BBH merger.

The era of GW astronomy is an exciting time for facilities like \Fermi, that excel at transient source discovery.  We have developed new pipelines and techniques to search the GBM and LAT data for transient sources, and set constraining upper limits using \Fermi data.  As LIGO and Virgo continue to become more sensitive, and new facilities come online (LIGO India, KAGRA), more BBH mergers will be detected, and BNS mergers (for which expectations for EM counterparts are much more concrete), are also expected to be observed.  This could finally identify the progenitors of sGRBs.

The GBM project is supported by NASA. Support for the German contribution to GBM was provided by the Bundesministerium f{\"u}r Bildung und Forschung (BMBF) via the Deutsches Zentrum f{\"u}r Luft und Raumfahrt (DLR) under contract number 50 QV 0301. 
AG is funded through the NASA Postdoctoral Fellowship Program.

The \textit{Fermi} LAT Collaboration acknowledges generous ongoing support from a number of agencies and institutes that have supported both the development and the operation of the LAT as well as scientific data analysis. These include the National Aeronautics and Space Administration and the Department of Energy in the United States, the Commissariat \`a l'Energie Atomique and the Centre National de la Recherche Scientifique / Institut National de Physique Nucl\'eaire et de Physique des Particules in France, the Agenzia Spaziale Italiana and the Istituto Nazionale di Fisica Nucleare in Italy, the Ministry of Education, Culture, Sports, Science and Technology (MEXT), High Energy Accelerator Research Organization (KEK) and Japan Aerospace Exploration Agency (JAXA) in Japan, and the K.~A.~Wallenberg Foundation, the Swedish Research Council and the Swedish National Space Board in Sweden.
 
Additional support for science analysis during the operations phase is gratefully acknowledged from the Istituto Nazionale di Astrofisica in Italy and the Centre National d'\'Etudes Spatiales in France.

NC and JB are supported by NSF grant PHY-1505373.

\bibliographystyle{yahapj}
\bibliography{references}

\begin{thebibliography}{}
\providecommand\natexlab[1]{#1}
\providecommand\JournalTitle[1]{#1}

\bibitem[{{Abbott} {et~al.}(2016{\natexlab{a}})}]{GW151226_LIGO}
{Abbott}, B.~P., {et~al.} 2016{\natexlab{a}}, \JournalTitle{PRL, in press}

\bibitem[{{Abbott} {et~al.}(2016{\natexlab{b}}){Abbott}, {Abbott}, {Abbott},
  {Abernathy}, {Acernese}, {Ackley}, {Adams}, {Adams}, \&
  et~al.}]{2016arXiv160203839T}
{Abbott}, B.~P., {Abbott}, R., {Abbott}, T.~D., {et~al.} 2016{\natexlab{b}},
  \JournalTitle{ArXiv e-prints},
  \href{http://arxiv.org/abs/1602.03839}{{\sffamily arXiv:1602.03839 [gr-qc]}}

\bibitem[{{Abbott} {et~al.}(2016{\natexlab{c}}){Abbott}, {Abbott}, {Abbott},
  {Abernathy}, {Acernese}, {Ackley}, {Adams}, {Adams}, {Addesso}, {Adhikari},
  \& et~al.}]{2016arXiv160208492A}
---. 2016{\natexlab{c}}, \JournalTitle{ArXiv e-prints},
  \href{http://arxiv.org/abs/1602.08492}{{\sffamily arXiv:1602.08492
  [astro-ph.HE]}}

\bibitem[{{Abbott} {et~al.}(2016{\natexlab{d}}){Abbott}, {Abbott}, {Abbott},
  {Abernathy}, {Acernese}, {Ackley}, {Adams}, {Adams}, {Addesso}, {Adhikari},
  \& et~al.}]{2016PhRvL.116f1102A}
---. 2016{\natexlab{d}},
  \href{http://dx.doi.org/10.1103/PhysRevLett.116.061102}{\JournalTitle{Physical
  Review Letters}, 116, 061102}

\bibitem[{{Abbott} {et~al.}(2016{\natexlab{e}}){Abbott}, {Abbott}, {Abbott},
  {Abernathy}, {Acernese}, {Ackley}, {Adams}, {Adams}, {Addesso}, {Adhikari},
  \& et~al.}]{2016LRR....19....1A}
---. 2016{\natexlab{e}},
  \href{http://dx.doi.org/10.1007/lrr-2016-1}{\JournalTitle{Living Reviews in
  Relativity}, 19}, \href{http://arxiv.org/abs/1304.0670}{{\sffamily
  arXiv:1304.0670 [gr-qc]}}

\bibitem[{{Abdo} {et~al.}(2009){Abdo}, {Ackermann}, {Ajello}, {Asano},
  {Atwood}, {Axelsson}, {Baldini}, {Ballet}, {Barbiellini}, {Baring}, \&
  et~al.}]{grb090510_nature}
{Abdo}, A.~A., {Ackermann}, M., {Ajello}, M., {et~al.} 2009,
  \href{http://dx.doi.org/10.1038/nature08574}{\JournalTitle{\nat}, 462, 331}

\bibitem[{{Acero} {et~al.}(2015){Acero}, {Ackermann}, {Ajello}, {Albert},
  {Atwood}, {Axelsson}, {Baldini}, {Ballet}, {Barbiellini}, {Bastieri},
  {Belfiore}, {Bellazzini}, {Bissaldi}, {Blandford}, {Bloom}, {Bogart},
  {Bonino}, {Bottacini}, {Bregeon}, {Britto}, {Bruel}, {Buehler}, {Burnett},
  {Buson}, {Caliandro}, {Cameron}, {Caputo}, {Caragiulo}, {Caraveo},
  {Casandjian}, {Cavazzuti}, {Charles}, {Chaves}, {Chekhtman}, {Cheung},
  {Chiang}, {Chiaro}, {Ciprini}, {Claus}, {Cohen-Tanugi}, {Cominsky}, {Conrad},
  {Cutini}, {D'Ammando}, {de Angelis}, {DeKlotz}, {de Palma}, {Desiante},
  {Digel}, {Di Venere}, {Drell}, {Dubois}, {Dumora}, {Favuzzi}, {Fegan},
  {Ferrara}, {Finke}, {Franckowiak}, {Fukazawa}, {Funk}, {Fusco}, {Gargano},
  {Gasparrini}, {Giebels}, {Giglietto}, {Giommi}, {Giordano}, {Giroletti},
  {Glanzman}, {Godfrey}, {Grenier}, {Grondin}, {Grove}, {Guillemot}, {Guiriec},
  {Hadasch}, {Harding}, {Hays}, {Hewitt}, {Hill}, {Horan}, {Iafrate}, {Jogler},
  {J{\'o}hannesson}, {Johnson}, {Johnson}, {Johnson}, {Johnson}, {Kamae},
  {Kataoka}, {Katsuta}, {Kuss}, {La Mura}, {Landriu}, {Larsson}, {Latronico},
  {Lemoine-Goumard}, {Li}, {Li}, {Longo}, {Loparco}, {Lott}, {Lovellette},
  {Lubrano}, {Madejski}, {Massaro}, {Mayer}, {Mazziotta}, {McEnery},
  {Michelson}, {Mirabal}, {Mizuno}, {Moiseev}, {Mongelli}, {Monzani},
  {Morselli}, {Moskalenko}, {Murgia}, {Nuss}, {Ohno}, {Ohsugi}, {Omodei},
  {Orienti}, {Orlando}, {Ormes}, {Paneque}, {Panetta}, {Perkins},
  {Pesce-Rollins}, {Piron}, {Pivato}, {Porter}, {Racusin}, {Rando}, {Razzano},
  {Razzaque}, {Reimer}, {Reimer}, {Reposeur}, {Rochester}, {Romani},
  {Salvetti}, {S{\'a}nchez-Conde}, {Saz Parkinson}, {Schulz}, {Siskind},
  {Smith}, {Spada}, {Spandre}, {Spinelli}, {Stephens}, {Strong}, {Suson},
  {Takahashi}, {Takahashi}, {Tanaka}, {Thayer}, {Thayer}, {Thompson},
  {Tibaldo}, {Tibolla}, {Torres}, {Torresi}, {Tosti}, {Troja}, {Van Klaveren},
  {Vianello}, {Winer}, {Wood}, {Wood}, {Zimmer}, \& {Fermi-LAT
  Collaboration}}]{3fgl}
{Acero}, F., {Ackermann}, M., {Ajello}, M., {et~al.} 2015,
  \href{http://dx.doi.org/10.1088/0067-0049/218/2/23}{\JournalTitle{\apjs},
  218, 23}

\bibitem[{{Acero} {et~al.}(2016)}]{2016ApJS..223...26A}
{Acero}, F., {et~al.} 2016,
  \href{http://dx.doi.org/10.3847/0067-0049/223/2/26}{\JournalTitle{\apjs},
  223, 26}

\bibitem[{{Ackermann} {et~al.}(2010){Ackermann}, {Asano}, {Atwood}, {Axelsson},
  {Baldini}, {Ballet}, {Barbiellini}, {Baring}, {Bastieri}, {Bechtol},
  {Bellazzini}, {Berenji}, {Bhat}, {Bissaldi}, {Blandford}, {Bloom},
  {Bonamente}, {Borgland}, {Bouvier}, {Bregeon}, {Brez}, {Briggs}, {Brigida},
  {Bruel}, {Buson}, {Caliandro}, {Cameron}, {Caraveo}, {Carrigan},
  {Casandjian}, {Cecchi}, {{\c C}elik}, {Charles}, {Chiang}, {Ciprini},
  {Claus}, {Cohen-Tanugi}, {Connaughton}, {Conrad}, {Dermer}, {de Palma},
  {Dingus}, {Silva}, {Drell}, {Dubois}, {Dumora}, {Farnier}, {Favuzzi},
  {Fegan}, {Finke}, {Focke}, {Frailis}, {Fukazawa}, {Fusco}, {Gargano},
  {Gasparrini}, {Gehrels}, {Germani}, {Giglietto}, {Giordano}, {Glanzman},
  {Godfrey}, {Granot}, {Grenier}, {Grondin}, {Grove}, {Guiriec}, {Hadasch},
  {Harding}, {Hays}, {Horan}, {Hughes}, {J{\'o}hannesson}, {Johnson}, {Kamae},
  {Katagiri}, {Kataoka}, {Kawai}, {Kippen}, {Kn{\"o}dlseder}, {Kocevski},
  {Kouveliotou}, {Kuss}, {Lande}, {Latronico}, {Lemoine-Goumard}, {Llena
  Garde}, {Longo}, {Loparco}, {Lott}, {Lovellette}, {Lubrano}, {Makeev},
  {Mazziotta}, {McEnery}, {McGlynn}, {Meegan}, {M{\'e}sz{\'a}ros}, {Michelson},
  {Mitthumsiri}, {Mizuno}, {Moiseev}, {Monte}, {Monzani}, {Moretti},
  {Morselli}, {Moskalenko}, {Murgia}, {Nakajima}, {Nakamori}, {Nolan},
  {Norris}, {Nuss}, {Ohno}, {Ohsugi}, {Omodei}, {Orlando}, {Ormes}, {Ozaki},
  {Paciesas}, {Paneque}, {Panetta}, {Parent}, {Pelassa}, {Pepe},
  {Pesce-Rollins}, {Piron}, {Preece}, {Rain{\`o}}, {Rando}, {Razzano},
  {Razzaque}, {Reimer}, {Ritz}, {Rodriguez}, {Roth}, {Ryde}, {Sadrozinski},
  {Sander}, {Scargle}, {Schalk}, {Sgr{\`o}}, {Siskind}, {Smith}, {Spandre},
  {Spinelli}, {Stamatikos}, {Stecker}, {Strickman}, {Suson}, {Tajima},
  {Takahashi}, {Takahashi}, {Tanaka}, {Thayer}, {Thayer}, {Thompson},
  {Tibaldo}, {Toma}, {Torres}, {Tosti}, {Tramacere}, {Uchiyama}, {Uehara},
  {Usher}, {van der Horst}, {Vasileiou}, {Vilchez}, {Vitale}, {von Kienlin},
  {Waite}, {Wang}, {Wilson-Hodge}, {Winer}, {Wu}, {Yamazaki}, {Yang}, {Ylinen},
  \& {Ziegler}}]{grb090510_lat}
{Ackermann}, M., {Asano}, K., {Atwood}, W.~B., {et~al.} 2010,
  \href{http://dx.doi.org/10.1088/0004-637X/716/2/1178}{\JournalTitle{\apj},
  716, 1178}

\bibitem[{{Ackermann} {et~al.}(2013{\natexlab{a}}){Ackermann}, {Ajello},
  {Asano}, {Baldini}, {Barbiellini}, {Baring}, {Bastieri}, {Bellazzini},
  {Blandford}, {Bonamente}, {Borgland}, {Bottacini}, {Bregeon}, {Brigida},
  {Bruel}, {Buehler}, {Buson}, {Caliandro}, {Cameron}, {Caraveo}, {Cecchi},
  {Charles}, {Chaves}, {Chekhtman}, {Chiang}, {Ciprini}, {Claus},
  {Cohen-Tanugi}, {Conrad}, {Cutini}, {D'Ammando}, {de Angelis}, {de Palma},
  {Dermer}, {Silva}, {Drell}, {Drlica-Wagner}, {Favuzzi}, {Fegan}, {Focke},
  {Franckowiak}, {Fukazawa}, {Fusco}, {Gargano}, {Gasparrini}, {Gehrels},
  {Giglietto}, {Giordano}, {Giroletti}, {Glanzman}, {Godfrey}, {Granot},
  {Greiner}, {Grenier}, {Grove}, {Guiriec}, {Hadasch}, {Hanabata}, {Hayashida},
  {Hays}, {Hughes}, {Jackson}, {Jogler}, {J{\'o}hannesson}, {Johnson},
  {Kn{\"o}dlseder}, {Kocevski}, {Kuss}, {Lande}, {Larsson}, {Latronico},
  {Longo}, {Loparco}, {Lovellette}, {Lubrano}, {Mazziotta}, {McEnery},
  {Mehault}, {M{\'e}sz{\'a}ros}, {Michelson}, {Mitthumsiri}, {Mizuno}, {Monte},
  {Monzani}, {Moretti}, {Morselli}, {Moskalenko}, {Murgia}, {Naumann-Godo},
  {Norris}, {Nuss}, {Nymark}, {Ohno}, {Ohsugi}, {Omodei}, {Orienti}, {Orlando},
  {Paneque}, {Perkins}, {Pesce-Rollins}, {Piron}, {Pivato}, {Racusin},
  {Rain{\`o}}, {Rando}, {Razzano}, {Razzaque}, {Reimer}, {Reimer}, {Romoli},
  {Roth}, {Ryde}, {Sanchez}, {Sgr{\`o}}, {Siskind}, {Sonbas}, {Spinelli},
  {Stamatikos}, {Takahashi}, {Tanaka}, {Thayer}, {Thayer}, {Tibaldo},
  {Tinivella}, {Tosti}, {Troja}, {Usher}, {Vandenbroucke}, {Vasileiou},
  {Vianello}, {Vitale}, {Waite}, {Winer}, {Wood}, {Yang}, {Gruber}, {Bhat},
  {Bissaldi}, {Briggs}, {Burgess}, {Connaughton}, {Foley}, {Kippen},
  {Kouveliotou}, {McBreen}, {McGlynn}, {Paciesas}, {Pelassa}, {Preece}, {Rau},
  {van der Horst}, {von Kienlin}, {Kann}, {Filgas}, {Klose}, {Kr{\"u}hler},
  {Fukui}, {Sako}, {Tristram}, {Oates}, {Ukwatta}, \&
  {Littlejohns}}]{grb110731}
{Ackermann}, M., {Ajello}, M., {Asano}, K., {et~al.} 2013{\natexlab{a}},
  \href{http://dx.doi.org/10.1088/0004-637X/763/2/71}{\JournalTitle{\apj}, 763,
  71}

\bibitem[{{Ackermann} {et~al.}(2013{\natexlab{b}}){Ackermann}, {Ajello},
  {Albert}, {Allafort}, {Antolini}, {Baldini}, {Ballet}, {Barbiellini},
  {Bastieri}, {Bechtol}, {Bellazzini}, {Blandford}, {Bloom}, {Bonamente},
  {Bottacini}, {Bouvier}, {Brandt}, {Bregeon}, {Brigida}, {Bruel}, {Buehler},
  {Buson}, {Caliandro}, {Cameron}, {Caraveo}, {Cavazzuti}, {Cecchi}, {Charles},
  {Chekhtman}, {Cheung}, {Chiang}, {Chiaro}, {Ciprini}, {Claus},
  {Cohen-Tanugi}, {Conrad}, {Cutini}, {Dalton}, {D'Ammando}, {de Angelis}, {de
  Palma}, {Dermer}, {Di Venere}, {Drell}, {Drlica-Wagner}, {Favuzzi}, {Fegan},
  {Ferrara}, {Focke}, {Franckowiak}, {Fukazawa}, {Funk}, {Fusco}, {Gargano},
  {Gasparrini}, {Germani}, {Giglietto}, {Giordano}, {Giroletti}, {Glanzman},
  {Godfrey}, {Grenier}, {Grondin}, {Grove}, {Guiriec}, {Hadasch}, {Hanabata},
  {Harding}, {Hayashida}, {Hays}, {Hewitt}, {Hill}, {Horan}, {Hou}, {Hughes},
  {Inoue}, {Jackson}, {Jogler}, {J{\'o}hannesson}, {Johnson}, {Kamae},
  {Kataoka}, {Kawano}, {Kn{\"o}dlseder}, {Kuss}, {Lande}, {Larsson},
  {Latronico}, {Lemoine-Goumard}, {Longo}, {Loparco}, {Lott}, {Lovellette},
  {Lubrano}, {Mayer}, {Mazziotta}, {McEnery}, {Michelson}, {Mitthumsiri},
  {Mizuno}, {Monte}, {Monzani}, {Morselli}, {Moskalenko}, {Murgia}, {Nemmen},
  {Nuss}, {Ohsugi}, {Okumura}, {Omodei}, {Orienti}, {Orlando}, {Ormes},
  {Paneque}, {Panetta}, {Perkins}, {Pesce-Rollins}, {Piron}, {Pivato},
  {Porter}, {Rain{\`o}}, {Rando}, {Razzano}, {Reimer}, {Reimer}, {Romoli},
  {Roth}, {S{\'a}nchez-Conde}, {Scargle}, {Schulz}, {Sgr{\`o}}, {Siskind},
  {Spandre}, {Spinelli}, {Suson}, {Takahashi}, {Takeuchi}, {Thayer}, {Thayer},
  {Thompson}, {Tibaldo}, {Tinivella}, {Torres}, {Tosti}, {Troja}, {Tronconi},
  {Usher}, {Vandenbroucke}, {Vasileiou}, {Vianello}, {Vitale}, {Winer}, {Wood},
  {Wood}, \& {Yang}}]{fava1}
{Ackermann}, M., {Ajello}, M., {Albert}, A., {et~al.} 2013{\natexlab{b}},
  \href{http://dx.doi.org/10.1088/0004-637X/771/1/57}{\JournalTitle{\apj}, 771,
  57}

\bibitem[{{Ackermann} {et~al.}(2013{\natexlab{c}}){Ackermann}, {Ajello},
  {Asano}, {Axelsson}, {Baldini}, {Ballet}, {Barbiellini}, {Bastieri},
  {Bechtol}, {Bellazzini}, {Bhat}, {Bissaldi}, {Bloom}, {Bonamente}, {Bonnell},
  {Bouvier}, {Brandt}, {Bregeon}, {Brigida}, {Bruel}, {Buehler}, {Burgess},
  {Buson}, {Byrne}, {Caliandro}, {Cameron}, {Caraveo}, {Cecchi}, {Charles},
  {Chaves}, {Chekhtman}, {Chiang}, {Chiaro}, {Ciprini}, {Claus},
  {Cohen-Tanugi}, {Connaughton}, {Conrad}, {Cutini}, {D'Ammando}, {de Angelis},
  {de Palma}, {Dermer}, {Desiante}, {Digel}, {Dingus}, {Di Venere}, {Drell},
  {Drlica-Wagner}, {Dubois}, {Favuzzi}, {Ferrara}, {Fitzpatrick}, {Foley},
  {Franckowiak}, {Fukazawa}, {Fusco}, {Gargano}, {Gasparrini}, {Gehrels},
  {Germani}, {Giglietto}, {Giommi}, {Giordano}, {Giroletti}, {Glanzman},
  {Godfrey}, {Goldstein}, {Granot}, {Grenier}, {Grove}, {Gruber}, {Guiriec},
  {Hadasch}, {Hanabata}, {Hayashida}, {Horan}, {Hou}, {Hughes}, {Inoue},
  {Jackson}, {Jogler}, {J{\'o}hannesson}, {Johnson}, {Johnson}, {Kamae},
  {Kataoka}, {Kawano}, {Kippen}, {Kn{\"o}dlseder}, {Kocevski}, {Kouveliotou},
  {Kuss}, {Lande}, {Larsson}, {Latronico}, {Lee}, {Longo}, {Loparco},
  {Lovellette}, {Lubrano}, {Massaro}, {Mayer}, {Mazziotta}, {McBreen},
  {McEnery}, {McGlynn}, {Michelson}, {Mizuno}, {Moiseev}, {Monte}, {Monzani},
  {Moretti}, {Morselli}, {Murgia}, {Nemmen}, {Nuss}, {Nymark}, {Ohno},
  {Ohsugi}, {Omodei}, {Orienti}, {Orlando}, {Paciesas}, {Paneque}, {Panetta},
  {Pelassa}, {Perkins}, {Pesce-Rollins}, {Piron}, {Pivato}, {Porter}, {Preece},
  {Racusin}, {Rain{\`o}}, {Rando}, {Rau}, {Razzano}, {Razzaque}, {Reimer},
  {Reimer}, {Reposeur}, {Ritz}, {Romoli}, {Roth}, {Ryde}, {Saz Parkinson},
  {Schalk}, {Sgr{\`o}}, {Siskind}, {Sonbas}, {Spandre}, {Spinelli}, {Suson},
  {Tajima}, {Takahashi}, {Takeuchi}, {Tanaka}, {Thayer}, {Thayer}, {Thompson},
  {Tibaldo}, {Tierney}, {Tinivella}, {Torres}, {Tosti}, {Troja}, {Tronconi},
  {Usher}, {Vandenbroucke}, {van der Horst}, {Vasileiou}, {Vianello}, {Vitale},
  {von Kienlin}, {Winer}, {Wood}, {Wood}, {Xiong}, \& {Yang}}]{latgrbcat1}
{Ackermann}, M., {Ajello}, M., {Asano}, K., {et~al.} 2013{\natexlab{c}},
  \href{http://dx.doi.org/10.1088/0067-0049/209/1/11}{\JournalTitle{\apjs},
  209, 11}

\bibitem[{{Ackermann} {et~al.}(2016){Ackermann}, {Ajello}, {Albert},
  {Anderson}, {Arimoto}, {Atwood}, {Axelsson}, {Baldini}, {Ballet},
  {Barbiellini}, {Baring}, {Bastieri}, {Becerra Gonzalez}, {Bellazzini},
  {Bissaldi}, {Blandford}, {Bloom}, {Bonino}, {Bottacini}, {Brandt}, {Bregeon},
  {Britto}, {Bruel}, {Buehler}, {Burnett}, {Buson}, {Caliandro}, {Cameron},
  {Caputo}, {Caragiulo}, {Caraveo}, {Casandjian}, {Cavazzuti}, {Charles},
  {Chekhtman}, {Chiang}, {Chiaro}, {Ciprini}, {Cohen-Tanugi}, {Cominsky},
  {Condon}, {Costanza}, {Cuoco}, {Cutini}, {DAmmando}, {de Palma}, {Desiante},
  {Digel}, {Di Lalla}, {Di Mauro}, {Di Venere}, {Dom{\'{\i}}nguez}, {Drell},
  {Dubois}, {Dumora}, {Favuzzi}, {Fegan}, {Ferrara}, {Franckowiak}, {Fukazawa},
  {Funk}, {Fusco}, {Gargano}, {Gasparrini}, {Gehrels}, {Giglietto}, {Giomi},
  {Giommi}, {Giordano}, {Giroletti}, {Glanzman}, {Godfrey}, {Gomez-Vargas},
  {Granot}, {Green}, {Grenier}, {Grondin}, {Grove}, {Guillemot}, {Guiriec},
  {Hadasch}, {Harding}, {Hays}, {Hewitt}, {Hill}, {Horan}, {Jogler},
  {J{\'o}hannesson}, {Kamae}, {Kensei}, {Kocevski}, {Kuss}, {La Mura},
  {Larsson}, {Latronico}, {Lemoine-Goumard}, {Li}, {Li}, {Longo}, {Loparco},
  {Lovellette}, {Lubrano}, {Madejski}, {Magill}, {Maldera}, {Manfreda},
  {Marelli}, {Mayer}, {Mazziotta}, {McEnery}, {Meyer}, {Michelson}, {Mirabal},
  {Mizuno}, {Moiseev}, {Monzani}, {Moretti}, {Morselli}, {Moskalenko},
  {Murgia}, {Negro}, {Nuss}, {Ohsugi}, {Omodei}, {Orienti}, {Orlando}, {Ormes},
  {Paneque}, {Perkins}, {Pesce-Rollins}, {Piron}, {Pivato}, {Porter},
  {Racusin}, {Rain{\`o}}, {Rando}, {Razzaque}, {Reimer}, {Reimer}, {Reposeur},
  {Ritz}, {Rochester}, {Romani}, {Saz Parkinson}, {Sgr{\`o}}, {Simone},
  {Siskind}, {Smith}, {Spada}, {Spandre}, {Spinelli}, {Suson}, {Tajima},
  {Thayer}, {Thayer}, {Thompson}, {Tibaldo}, {Torres}, {Troja}, {Uchiyama},
  {Venters}, {Vianello}, {Wood}, {Wood}, {Zaharijas}, {Zhu}, \&
  {Zimmer}}]{GW150914_LAT}
{Ackermann}, M., {Ajello}, M., {Albert}, A., {et~al.} 2016,
  \href{http://dx.doi.org/10.3847/2041-8205/823/1/L2}{\JournalTitle{\apjl},
  823, L2}

\bibitem[{{Atwood} {et~al.}(2009){Atwood}, {Abdo}, {Ackermann}, {Althouse},
  {Anderson}, {Axelsson}, {Baldini}, {Ballet}, {Band}, {Barbiellini}, \&
  et~al.}]{2009ApJ...697.1071A}
{Atwood}, W.~B., {Abdo}, A.~A., {Ackermann}, M., {et~al.} 2009,
  \href{http://dx.doi.org/10.1088/0004-637X/697/2/1071}{\JournalTitle{\apj},
  697, 1071}

\bibitem[{{Bagoly} {et~al.}(2016){Bagoly}, {Sz{\'e}csi}, {Bal{\'a}zs},
  {Csabai}, {Horv{\'a}th}, {Dobos}, {Lichtenberger}, \&
  {T{\'o}th}}]{2016arXiv160306611B}
{Bagoly}, Z., {Sz{\'e}csi}, D., {Bal{\'a}zs}, L.~G., {et~al.} 2016,
  \JournalTitle{ArXiv e-prints},
  \href{http://arxiv.org/abs/1603.06611}{{\sffamily arXiv:1603.06611
  [astro-ph.HE]}}

\bibitem[{{Bhat} {et~al.}(2016){Bhat}, {Meegan}, {von Kienlin}, {Paciesas},
  {Briggs}, {Burgess}, {Burns}, {Chaplin}, {Cleveland}, {Collazzi},
  {Connaughton}, {Diekmann}, {Fitzpatrick}, {Gibby}, {Giles}, {Goldstein},
  {Greiner}, {Jenke}, {Kippen}, {Kouveliotou}, {Mailyan}, {McBreen}, {Pelassa},
  {Preece}, {Roberts}, {Sparke}, {Stanbro}, {Veres}, {Wilson-Hodge}, {Xiong},
  {Younes}, {Yu}, \& {Zhang}}]{3rdgbm}
{Bhat}, P., {Meegan}, C.~A., {von Kienlin}, A., {et~al.} 2016,
  \href{http://dx.doi.org/10.3847/0067-0049/223/2/28}{\JournalTitle{\apjs},
  223, 28}

\bibitem[{{Blackburn} {et~al.}(2015){Blackburn}, {Briggs}, {Camp},
  {Christensen}, {Connaughton}, {Jenke}, {Remillard}, \&
  {Veitch}}]{2015ApJS..217....8B}
{Blackburn}, L., {Briggs}, M.~S., {Camp}, J., {et~al.} 2015,
  \href{http://dx.doi.org/10.1088/0067-0049/217/1/8}{\JournalTitle{\apjs}, 217,
  8}

\bibitem[{{Blandford} \& {Znajek}(1977)}]{1977MNRAS.179..433B}
{Blandford}, R.~D., \& {Znajek}, R.~L. 1977,
  \href{http://dx.doi.org/10.1093/mnras/179.3.433}{\JournalTitle{\mnras}, 179,
  433}

\bibitem[{{Branchina} \& {De Domenico}(2016)}]{2016arXiv160408530B}
{Branchina}, V., \& {De Domenico}, M. 2016, \JournalTitle{ArXiv e-prints},
  \href{http://arxiv.org/abs/1604.08530}{{\sffamily arXiv:1604.08530 [gr-qc]}}

\bibitem[{{Ciolfi} \& {Siegel}(2015)}]{Ciolfi2015}
{Ciolfi}, R., \& {Siegel}, D.~M. 2015, \JournalTitle{ArXiv e-prints},
  \href{http://arxiv.org/abs/1505.01420}{{\sffamily arXiv:1505.01420
  [astro-ph.HE]}}

\bibitem[{{Connaughton} {et~al.}(2015){Connaughton}, {Briggs}, {Goldstein},
  {Meegan}, {Paciesas}, {Preece}, {Wilson-Hodge}, {Gibby}, {Greiner}, {Gruber},
  {Jenke}, {Kippen}, {Pelassa}, {Xiong}, {Yu}, {Bhat}, {Burgess}, {Byrne},
  {Fitzpatrick}, {Foley}, {Giles}, {Guiriec}, {van der Horst}, {von Kienlin},
  {McBreen}, {McGlynn}, {Tierney}, \& {Zhang}}]{2015ApJS..216...32C}
{Connaughton}, V., {Briggs}, M.~S., {Goldstein}, A., {et~al.} 2015,
  \href{http://dx.doi.org/10.1088/0067-0049/216/2/32}{\JournalTitle{\apjs},
  216, 32}

\bibitem[{{Connaughton} {et~al.}(2016){Connaughton}, {Burns}, {Goldstein},
  {Briggs}, {Zhang}, {Hui}, {Jenke}, {Racusin}, {Wilson-Hodge}, {Bhat},
  {Bissaldi}, {Cleveland}, {Fitzpatrick}, {Giles}, {Gibby}, {Greiner}, {von
  Kienlin}, {Kippen}, {McBreen}, {Mailyan}, {Meegan}, {Paciesas}, {Preece},
  {Roberts}, {Sparke}, {Stanbro}, {Toelge}, {Veres}, {Yu}, \&
  {authors}}]{GW150914_GBM}
{Connaughton}, V., {Burns}, E., {Goldstein}, A., {et~al.} 2016,
  \JournalTitle{\apjl, in press},
  \href{http://arxiv.org/abs/1602.03920}{{\sffamily arXiv:1602.03920
  [astro-ph.HE]}}

\bibitem[{{D'Avanzo} {et~al.}(2014){D'Avanzo}, {Salvaterra}, {Bernardini},
  {Nava}, {Campana}, {Covino}, {D'Elia}, {Ghirlanda}, {Ghisellini}, {Melandri},
  {Sbarufatti}, {Vergani}, \& {Tagliaferri}}]{2014MNRAS.442.2342D}
{D'Avanzo}, P., {Salvaterra}, R., {Bernardini}, M.~G., {et~al.} 2014,
  \href{http://dx.doi.org/10.1093/mnras/stu994}{\JournalTitle{\mnras}, 442,
  2342}

\bibitem[{{De Pasquale} {et~al.}(2010){De Pasquale}, {Schady}, {Kuin}, {Page},
  {Curran}, {Zane}, {Oates}, {Holland}, {Breeveld}, {Hoversten}, {Chincarini},
  {Grupe}, {Abdo}, {Ackermann}, {Ajello}, {Axelsson}, {Baldini}, {Ballet},
  {Barbiellini}, {Baring}, {Bastieri}, {Bechtol}, {Bellazzini}, {Berenji},
  {Bissaldi}, {Blandford}, {Bloom}, {Bonamente}, {Borgland}, {Bouvier},
  {Bregeon}, {Brez}, {Briggs}, {Brigida}, {Bruel}, {Burnett}, {Buson},
  {Caliandro}, {Cameron}, {Caraveo}, {Carrigan}, {Casandjian}, {Cecchi}, {{\c
  C}elik}, {Chekhtman}, {Chiang}, {Ciprini}, {Claus}, {Cohen-Tanugi},
  {Connaughton}, {Conrad}, {Dermer}, {de Angelis}, {de Palma}, {Dingus},
  {Silva}, {Drell}, {Dubois}, {Dumora}, {Farnier}, {Favuzzi}, {Fegan},
  {Fishman}, {Focke}, {Frailis}, {Fukazawa}, {Funk}, {Fusco}, {Gargano},
  {Gasparrini}, {Gehrels}, {Germani}, {Giglietto}, {Giordano}, {Glanzman},
  {Godfrey}, {Granot}, {Greiner}, {Grenier}, {Grove}, {Guillemot}, {Guiriec},
  {Harding}, {Hayashida}, {Hays}, {Horan}, {Hughes}, {Jackson},
  {J{\'o}hannesson}, {Johnson}, {Johnson}, {Kamae}, {Katagiri}, {Kataoka},
  {Kawai}, {Kerr}, {Kippen}, {Kn{\"o}dlseder}, {Kocevski}, {Kuss}, {Lande},
  {Latronico}, {Lemoine-Goumard}, {Longo}, {Loparco}, {Lott}, {Lovellette},
  {Lubrano}, {Makeev}, {Mazziotta}, {McEnery}, {McGlynn}, {Meegan},
  {M{\'e}sz{\'a}ros}, {Meurer}, {Michelson}, {Mitthumsiri}, {Mizuno}, {Monte},
  {Monzani}, {Moretti}, {Morselli}, {Moskalenko}, {Murgia}, {Nolan}, {Norris},
  {Nuss}, {Ohno}, {Ohsugi}, {Omodei}, {Orlando}, {Ormes}, {Paciesas},
  {Paneque}, {Panetta}, {Parent}, {Pelassa}, {Pepe}, {Pesce-Rollins}, {Piron},
  {Porter}, {Preece}, {Rain{\`o}}, {Rando}, {Razzano}, {Reimer}, {Reimer},
  {Reposeur}, {Ritz}, {Rochester}, {Rodriguez}, {Roth}, {Ryde}, {Sadrozinski},
  {Sander}, {Saz Parkinson}, {Scargle}, {Schalk}, {Sgr{\`o}}, {Siskind},
  {Smith}, {Spandre}, {Spinelli}, {Stamatikos}, {Starck}, {Stecker},
  {Strickman}, {Suson}, {Tajima}, {Takahashi}, {Tanaka}, {Thayer}, {Thayer},
  {Thompson}, {Tibaldo}, {Toma}, {Torres}, {Tosti}, {Tramacere}, {Uchiyama},
  {Uehara}, {Usher}, {van der Horst}, {Vasileiou}, {Vilchez}, {Vitale}, {von
  Kienlin}, {Waite}, {Wang}, {Winer}, {Wood}, {Wu}, {Yamazaki}, {Ylinen}, \&
  {Ziegler}}]{grb090510}
{De Pasquale}, M., {Schady}, P., {Kuin}, N.~P.~M., {et~al.} 2010,
  \href{http://dx.doi.org/10.1088/2041-8205/709/2/L146}{\JournalTitle{\apjl},
  709, L146}

\bibitem[{{Eichler} {et~al.}(1989){Eichler}, {Livio}, {Piran}, \&
  {Schramm}}]{Eichler1989}
{Eichler}, D., {Livio}, M., {Piran}, T., \& {Schramm}, D.~N. 1989,
  \href{http://dx.doi.org/10.1038/340126a0}{\JournalTitle{\nat}, 340, 126}

\bibitem[{{Fermi-LAT Collaboration}(2016)}]{latgrbcat2}
{Fermi-LAT Collaboration}. 2016, \JournalTitle{in preparation}

\bibitem[{{Fong} {et~al.}(2015){Fong}, {Berger}, {Margutti}, \&
  {Zauderer}}]{fong15}
{Fong}, W., {Berger}, E., {Margutti}, R., \& {Zauderer}, B.~A. 2015,
  \href{http://dx.doi.org/10.1088/0004-637X/815/2/102}{\JournalTitle{\apj},
  815, 102}

\bibitem[{{Fraschetti}(2016)}]{Fraschetti2016}
{Fraschetti}, F. 2016, \JournalTitle{ArXiv e-prints},
  \href{http://arxiv.org/abs/1603.01950}{{\sffamily arXiv:1603.01950
  [astro-ph.HE]}}

\bibitem[{{Goldstein} {et~al.}(2012){Goldstein}, {Burgess}, {Preece}, {Briggs},
  {Guiriec}, {van der Horst}, {Connaughton}, {Wilson-Hodge}, {Paciesas},
  {Meegan}, {von Kienlin}, {Bhat}, {Bissaldi}, {Chaplin}, {Diehl}, {Fishman},
  {Fitzpatrick}, {Foley}, {Gibby}, {Giles}, {Greiner}, {Gruber}, {Kippen},
  {Kouveliotou}, {McBreen}, {McGlynn}, {Rau}, \& {Tierney}}]{goldstein12}
{Goldstein}, A., {Burgess}, J.~M., {Preece}, R.~D., {et~al.} 2012,
  \href{http://dx.doi.org/10.1088/0067-0049/199/1/19}{\JournalTitle{\apjs},
  199, 19}

\bibitem[{{G{\'o}rski} {et~al.}(2005){G{\'o}rski}, {Hivon}, {Banday},
  {Wandelt}, {Hansen}, {Reinecke}, \& {Bartelmann}}]{HEALPix}
{G{\'o}rski}, K.~M., {Hivon}, E., {Banday}, A.~J., {et~al.} 2005,
  \href{http://dx.doi.org/10.1086/427976}{\JournalTitle{\apj}, 622, 759}

\bibitem[{{Greiner} {et~al.}(2016){Greiner}, {Burgess}, {Savchenko}, \&
  {Yu}}]{greiner16}
{Greiner}, J., {Burgess}, J.~M., {Savchenko}, V., \& {Yu}, H.-F. 2016,
  \JournalTitle{\apjl, in press},
  \href{http://arxiv.org/abs/1606.00314}{{\sffamily arXiv:1606.00314
  [astro-ph.HE]}}

\bibitem[{{Gruber} {et~al.}(2014){Gruber}, {Goldstein}, {Weller von Ahlefeld},
  {Narayana Bhat}, {Bissaldi}, {Briggs}, {Byrne}, {Cleveland}, {Connaughton},
  {Diehl}, {Fishman}, {Fitzpatrick}, {Foley}, {Gibby}, {Giles}, {Greiner},
  {Guiriec}, {van der Horst}, {von Kienlin}, {Kouveliotou}, {Layden}, {Lin},
  {Meegan}, {McGlynn}, {Paciesas}, {Pelassa}, {Preece}, {Rau}, {Wilson-Hodge},
  {Xiong}, {Younes}, \& {Yu}}]{gruber14}
{Gruber}, D., {Goldstein}, A., {Weller von Ahlefeld}, V., {et~al.} 2014,
  \href{http://dx.doi.org/10.1088/0067-0049/211/1/12}{\JournalTitle{\apjs},
  211, 12}

\bibitem[{{Janiuk} {et~al.}(2016){Janiuk}, {Bejger}, {Charzynski}, \&
  {Sukova}}]{Janiuk2016}
{Janiuk}, A., {Bejger}, M., {Charzynski}, S., \& {Sukova}, P. 2016,
  \JournalTitle{ArXiv e-prints},
  \href{http://arxiv.org/abs/1604.07132}{{\sffamily arXiv:1604.07132
  [astro-ph.HE]}}

\bibitem[{{Kelley} {et~al.}(2013){Kelley}, {Mandel}, \&
  {Ramirez-Ruiz}}]{2013PhRvD..87l3004K}
{Kelley}, L.~Z., {Mandel}, I., \& {Ramirez-Ruiz}, E. 2013,
  \href{http://dx.doi.org/10.1103/PhysRevD.87.123004}{\JournalTitle{\prd}, 87,
  123004}

\bibitem[{{Kouveliotou} {et~al.}(2013){Kouveliotou}, {Granot}, {Racusin},
  {Bellm}, {Vianello}, {Oates}, {Fryer}, {Boggs}, {Christensen}, {Craig},
  {Dermer}, {Gehrels}, {Hailey}, {Harrison}, {Melandri}, {McEnery}, {Mundell},
  {Stern}, {Tagliaferri}, \& {Zhang}}]{kouveliotou13}
{Kouveliotou}, C., {Granot}, J., {Racusin}, J.~L., {et~al.} 2013,
  \href{http://dx.doi.org/10.1088/2041-8205/779/1/L1}{\JournalTitle{\apjl},
  779, L1}

\bibitem[{{Lee} \& {Ramirez-Ruiz}(2007)}]{LeeRamirezRuiz2007}
{Lee}, W.~H., \& {Ramirez-Ruiz}, E. 2007,
  \href{http://dx.doi.org/10.1088/1367-2630/9/1/017}{\JournalTitle{New Journal
  of Physics}, 9, 17}

\bibitem[{{Loeb}(2016)}]{Loeb2016}
{Loeb}, A. 2016,
  \href{http://dx.doi.org/10.3847/2041-8205/819/2/L21}{\JournalTitle{\apjl},
  819, L21}

\bibitem[{{LVC}(2015)}]{GW151226_GCN}
{LVC}. 2015, \JournalTitle{GRB Coordinates Network}, 18728

\bibitem[{{LVC}(2016)}]{LVT151012_GCN}
---. 2016, \JournalTitle{GRB Coordinates Network}, 19341

\bibitem[{{Lyutikov}(2016)}]{Lyutikov2016}
{Lyutikov}, M. 2016, \JournalTitle{ArXiv e-prints},
  \href{http://arxiv.org/abs/1602.07352}{{\sffamily arXiv:1602.07352
  [astro-ph.HE]}}

\bibitem[{{Meegan} {et~al.}(2009){Meegan}, {Lichti}, {Bhat}, {Bissaldi},
  {Greiner}, {Hoover}, {van der Horst}, {von Kienlin}, {Kippen}, {Kouveliotou},
  {McBreen}, {Paciesas}, {Preece}, {Steinle}, {Wallace}, {Wilson}, \&
  {Wilson-Hodge}}]{2009ApJ...702..791M}
{Meegan}, C., {Lichti}, G., {Bhat}, P.~N., {et~al.} 2009,
  \href{http://dx.doi.org/10.1088/0004-637X/702/1/791}{\JournalTitle{\apj},
  702, 791}

\bibitem[{{Messick} {et~al.}(2016){Messick}, {Blackburn}, {Brady}, {Brockill},
  {Cannon}, {Caudill}, {Chamberlin}, {Creighton}, {Everett}, {Hanna}, {Lang},
  {Li}, {Meacher}, {Pankow}, {Privitera}, {Qi}, {Sachdev}, {Sadeghian},
  {Sathyaprakash}, {Singer}, {Thomas}, {Wade}, {Wade}, \&
  {Weinstein}}]{2016arXiv160404324M}
{Messick}, C., {Blackburn}, K., {Brady}, P., {et~al.} 2016, \JournalTitle{ArXiv
  e-prints}, \href{http://arxiv.org/abs/1604.04324}{{\sffamily arXiv:1604.04324
  [astro-ph.IM]}}

\bibitem[{{Metzger} \& {Berger}(2012)}]{2012ApJ...746...48M}
{Metzger}, B.~D., \& {Berger}, E. 2012,
  \href{http://dx.doi.org/10.1088/0004-637X/746/1/48}{\JournalTitle{\apj}, 746,
  48}

\bibitem[{{Moharana} {et~al.}(2016){Moharana}, {Razzaque}, {Gupta}, \&
  {Meszaros}}]{2016arXiv160208436M}
{Moharana}, R., {Razzaque}, S., {Gupta}, N., \& {Meszaros}, P. 2016,
  \JournalTitle{\prd, in press},
  \href{http://arxiv.org/abs/1602.08436}{{\sffamily arXiv:1602.08436
  [astro-ph.HE]}}

\bibitem[{{Murase} {et~al.}(2016){Murase}, {Kashiyama}, {M{\'e}sz{\'a}ros},
  {Shoemaker}, \& {Senno}}]{Murase2016}
{Murase}, K., {Kashiyama}, K., {M{\'e}sz{\'a}ros}, P., {Shoemaker}, I., \&
  {Senno}, N. 2016,
  \href{http://dx.doi.org/10.3847/2041-8205/822/1/L9}{\JournalTitle{\apjl},
  822, L9}

\bibitem[{{Nakar}(2007)}]{Nakar2007}
{Nakar}, E. 2007,
  \href{http://dx.doi.org/10.1016/j.physrep.2007.02.005}{\JournalTitle{\physrep},
  442, 166}

\bibitem[{{Narayan} {et~al.}(1992){Narayan}, {Paczynski}, \&
  {Piran}}]{Narayan1992}
{Narayan}, R., {Paczynski}, B., \& {Piran}, T. 1992,
  \href{http://dx.doi.org/10.1086/186493}{\JournalTitle{\apjl}, 395, L83}

\bibitem[{{Neyman} \& {Pearson}(1928)}]{Neyman1928}
{Neyman}, J., \& {Pearson}, E.~S. 1928,
  \href{http://dx.doi.org/10.1007/s11214-005-5097-2}{\JournalTitle{Biometrika},
  20, 175}

\bibitem[{{Nissanke} {et~al.}(2013){Nissanke}, {Kasliwal}, \&
  {Georgieva}}]{2013ApJ...767..124N}
{Nissanke}, S., {Kasliwal}, M., \& {Georgieva}, A. 2013,
  \href{http://dx.doi.org/10.1088/0004-637X/767/2/124}{\JournalTitle{\apj},
  767, 124}

\bibitem[{Olive {et~al.}(2014)}]{Agashe:2014kda}
Olive, K.~A., {et~al.} 2014,
  \href{http://dx.doi.org/10.1088/1674-1137/38/9/090001}{\JournalTitle{Chin.
  Phys.}, C38, 090001}

\bibitem[{{Perna} {et~al.}(2016){Perna}, {Lazzati}, \&
  {Giacomazzo}}]{2016ApJ...821L..18P}
{Perna}, R., {Lazzati}, D., \& {Giacomazzo}, B. 2016,
  \href{http://dx.doi.org/10.3847/2041-8205/821/1/L18}{\JournalTitle{\apjl},
  821, L18}

\bibitem[{{Rezzolla} {et~al.}(2011){Rezzolla}, {Giacomazzo}, {Baiotti},
  {Granot}, {Kouveliotou}, \& {Aloy}}]{Rezzolla2011}
{Rezzolla}, L., {Giacomazzo}, B., {Baiotti}, L., {et~al.} 2011,
  \href{http://dx.doi.org/10.1088/2041-8205/732/1/L6}{\JournalTitle{\apjl},
  732, L6}

\bibitem[{{Rosswog}(2005)}]{2005ApJ...634.1202R}
{Rosswog}, S. 2005,
  \href{http://dx.doi.org/10.1086/497062}{\JournalTitle{\apj}, 634, 1202}

\bibitem[{{Savchenko} {et~al.}(2016){Savchenko}, {Ferrigno}, {Mereghetti},
  {Natalucci}, {Bazzano}, {Bozzo}, {Brandt}, {Courvoisier}, {Diehl}, {Hanlon},
  {von Kienlin}, {Kuulkers}, {Laurent}, {Lebrun}, {Roques}, {Ubertini}, \&
  {Weidenspointner}}]{GW150914_integral}
{Savchenko}, V., {Ferrigno}, C., {Mereghetti}, S., {et~al.} 2016,
  \href{http://dx.doi.org/10.3847/2041-8205/820/2/L36}{\JournalTitle{\apjl},
  820, L36}

\bibitem[{{Tavani} {et~al.}(2016){Tavani}, {Pittori}, {Verrecchia},
  {Bulgarelli}, {Giuliani}, {Donnarumma}, {Argan}, {Trois}, {Lucarelli},
  {Marisaldi}, {Del Monte}, {Evangelista}, {Fioretti}, {Zoli}, {Piano},
  {Munar-Adrover}, {Antonelli}, {Barbiellini}, {Caraveo}, {Cattaneo}, {Costa},
  {Feroci}, {Ferrari}, {Longo}, {Mereghetti}, {Minervini}, {Morselli},
  {Pacciani}, {Pellizzoni}, {Picozza}, {Pilia}, {Rappoldi}, {Sabatini},
  {Vercellone}, {Vittorini}, {Giommi}, {Colafrancesco}, \&
  {Cardillo}}]{GW150914_AGILE}
{Tavani}, M., {Pittori}, C., {Verrecchia}, F., {et~al.} 2016,
  \JournalTitle{ArXiv e-prints},
  \href{http://arxiv.org/abs/1604.00955}{{\sffamily arXiv:1604.00955
  [astro-ph.HE]}}

\bibitem[{{Troja} {et~al.}(2008){Troja}, {King}, {O'Brien}, {Lyons}, \&
  {Cusumano}}]{2008MNRAS.385L..10T}
{Troja}, E., {King}, A.~R., {O'Brien}, P.~T., {Lyons}, N., \& {Cusumano}, G.
  2008,
  \href{http://dx.doi.org/10.1111/j.1745-3933.2007.00421.x}{\JournalTitle{\mnras},
  385, L10}

\bibitem[{{Troja} {et~al.}(2010){Troja}, {Rosswog}, \&
  {Gehrels}}]{2010ApJ...723.1711T}
{Troja}, E., {Rosswog}, S., \& {Gehrels}, N. 2010,
  \href{http://dx.doi.org/10.1088/0004-637X/723/2/1711}{\JournalTitle{\apj},
  723, 1711}

\bibitem[{{Usman} {et~al.}(2015){Usman}, {Nitz}, {Harry}, {Biwer}, {Brown},
  {Cabero}, {Capano}, {Dal Canton}, {Dent}, {Fairhurst}, {Kehl}, {Keppel},
  {Krishnan}, {Lenon}, {Lundgren}, {Nielsen}, {Pekowsky}, {Pfeiffer},
  {Saulson}, {West}, \& {Willis}}]{2015arXiv150802357U}
{Usman}, S.~A., {Nitz}, A.~H., {Harry}, I.~W., {et~al.} 2015,
  \JournalTitle{ArXiv e-prints},
  \href{http://arxiv.org/abs/1508.02357}{{\sffamily arXiv:1508.02357 [gr-qc]}}

\bibitem[{{Vasileiou} {et~al.}(2013){Vasileiou}, {Jacholkowska}, {Piron},
  {Bolmont}, {Couturier}, {Granot}, {Stecker}, {Cohen-Tanugi}, \&
  {Longo}}]{Vasileiou2013}
{Vasileiou}, V., {Jacholkowska}, A., {Piron}, F., {et~al.} 2013,
  \href{http://dx.doi.org/10.1103/PhysRevD.87.122001}{\JournalTitle{\prd}, 87,
  122001}

\bibitem[{{Veitch} {et~al.}(2015){Veitch}, {Raymond}, {Farr}, {Farr}, {Graff},
  {Vitale}, {Aylott}, {Blackburn}, {Christensen}, {Coughlin}, {Del Pozzo},
  {Feroz}, {Gair}, {Haster}, {Kalogera}, {Littenberg}, {Mandel},
  {O'Shaughnessy}, {Pitkin}, {Rodriguez}, {R{\"o}ver}, {Sidery}, {Smith}, {Van
  Der Sluys}, {Vecchio}, {Vousden}, \& {Wade}}]{2015PhRvD..91d2003V}
{Veitch}, J., {Raymond}, V., {Farr}, B., {et~al.} 2015,
  \href{http://dx.doi.org/10.1103/PhysRevD.91.042003}{\JournalTitle{\prd}, 91,
  042003}

\bibitem[{{Wilson-Hodge} {et~al.}(2012){Wilson-Hodge}, {Case}, {Cherry},
  {Rodi}, {Camero-Arranz}, {Jenke}, {Chaplin}, {Beklen}, {Finger}, {Bhat},
  {Briggs}, {Connaughton}, {Greiner}, {Kippen}, {Meegan}, {Paciesas}, {Preece},
  \& {von Kienlin}}]{2012ApJS..201...33W}
{Wilson-Hodge}, C.~A., {Case}, G.~L., {Cherry}, M.~L., {et~al.} 2012,
  \href{http://dx.doi.org/10.1088/0067-0049/201/2/33}{\JournalTitle{\apjs},
  201, 33}

\bibitem[{{Woosley}(2016)}]{2016arXiv160300511W}
{Woosley}, S.~E. 2016,
  \href{http://dx.doi.org/10.3847/2041-8205/824/1/L10}{\JournalTitle{\apjl},
  824, L10}

\bibitem[{{Yunes} {et~al.}(2016){Yunes}, {Yagi}, \&
  {Pretorius}}]{2016arXiv160308955Y}
{Yunes}, N., {Yagi}, K., \& {Pretorius}, F. 2016, \JournalTitle{\prd,
  submitted}, \href{http://arxiv.org/abs/1603.08955}{{\sffamily
  arXiv:1603.08955 [gr-qc]}}

\end{thebibliography}


\end{document}